# GridSim: A Toolkit for the Modeling and Simulation of Distributed Resource Management and Scheduling for Grid Computing

Rajkumar Buyya and Manzur Murshed[1]

School of Computer Science and Software Eng.
Monash University, Caulfield Campus
Melbourne, Vic 3145, Australia
*rajkumar@buyya.com*

Gippsland School of Computing and IT
Monash University, Gippsland Campus
Churchill, Vic 3842, Australia
*Manzur.Murshed@infotech.monash.edu.au*

**Abstract:** Clusters, grids, and peer-to-peer (P2P) networks have emerged as popular paradigms for next generation parallel and distributed computing. They enable aggregation of distributed resources for solving large-scale problems in science, engineering, and commerce. In grid and P2P computing environments, the resources are usually geographically distributed in *multiple administrative domains*, managed and owned by different organizations with different policies, and interconnected by wide-area networks or the Internet. This introduces a number of resource management and application scheduling challenges in the domain of security, resource and policy heterogeneity, fault tolerance, continuously changing resource conditions, and policies. The resource management and scheduling systems for grid computing need to manage resources and application execution depending on either resource consumers' or owners' requirements, and continuously adapt to changes in resource availability.

The management of resources and scheduling of applications in such large-scale distributed systems is a complex undertaking. In order to prove the effectiveness of resource brokers and associated scheduling algorithms, their performance needs to be evaluated under different scenarios such as varying number of resources and users with different requirements. In a grid environment, it is hard and even impossible to perform scheduler performance evaluation in a *repeatable* and *controllable* manner as resources and users are distributed across multiple organizations with their own policies. To overcome this limitation, we have developed a Java-based discrete-event grid simulation toolkit called *GridSim*. The toolkit supports modeling and simulation of heterogeneous grid resources (both time- and space-shared), users and application models. It provides primitives for creation of application tasks, mapping of tasks to resources, and their management. To demonstrate suitability of the GridSim toolkit, we have simulated a Nimrod-G like grid resource broker and evaluated the performance of deadline and budget constrained cost- and time-minimization scheduling algorithms.

## 1 Introduction

The proliferation of the Internet and the availability of powerful computers and high-speed networks as low-cost commodity components are changing the way we do large-scale parallel and distributed computing. The interest in coupling geographically distributed (computational) resources is also growing for solving large-scale problems, leading to what is popularly called the grid [3] and peer-to-peer (P2P) computing [4] networks. These enable sharing, selection and aggregation of suitable computational and data resources for solving large-scale data intensive problems in science, engineering, and commerce. A generic view of grid computing environment is shown in Figure 1. The grid consists of four key layers of components: fabric, core middleware, user-level middleware, and applications [6]. The grid fabric includes computers (low-end and high-end computers including clusters), networks, scientific instruments, and their resource management systems. The core grid middleware provides services that are essential for securely accessing remote resources uniformly and transparently. The services they provide include security and access management, remote job submission, storage, and resource information. The user-level middleware provides higher-level tools such as resource brokers, application development and adaptive runtime environment. The grid applications include those constructed using grid libraries or legacy applications that can be grid enabled using user-level middleware tools.

[1] Authors listed in surname order (to signify their similar effort!)

The user essentially interacts with a resource broker that hides the complexities of grid computing [7][8]. The broker discovers resources that the user can access using information services, negotiates for access costs using trading services, maps tasks to resources (scheduling), stages the application and data for processing (deployment), starts job execution, and finally gathers the results. It is also responsible for monitoring and tracking application execution progress along with adapting to the changes in grid runtime environment conditions and resource failures.

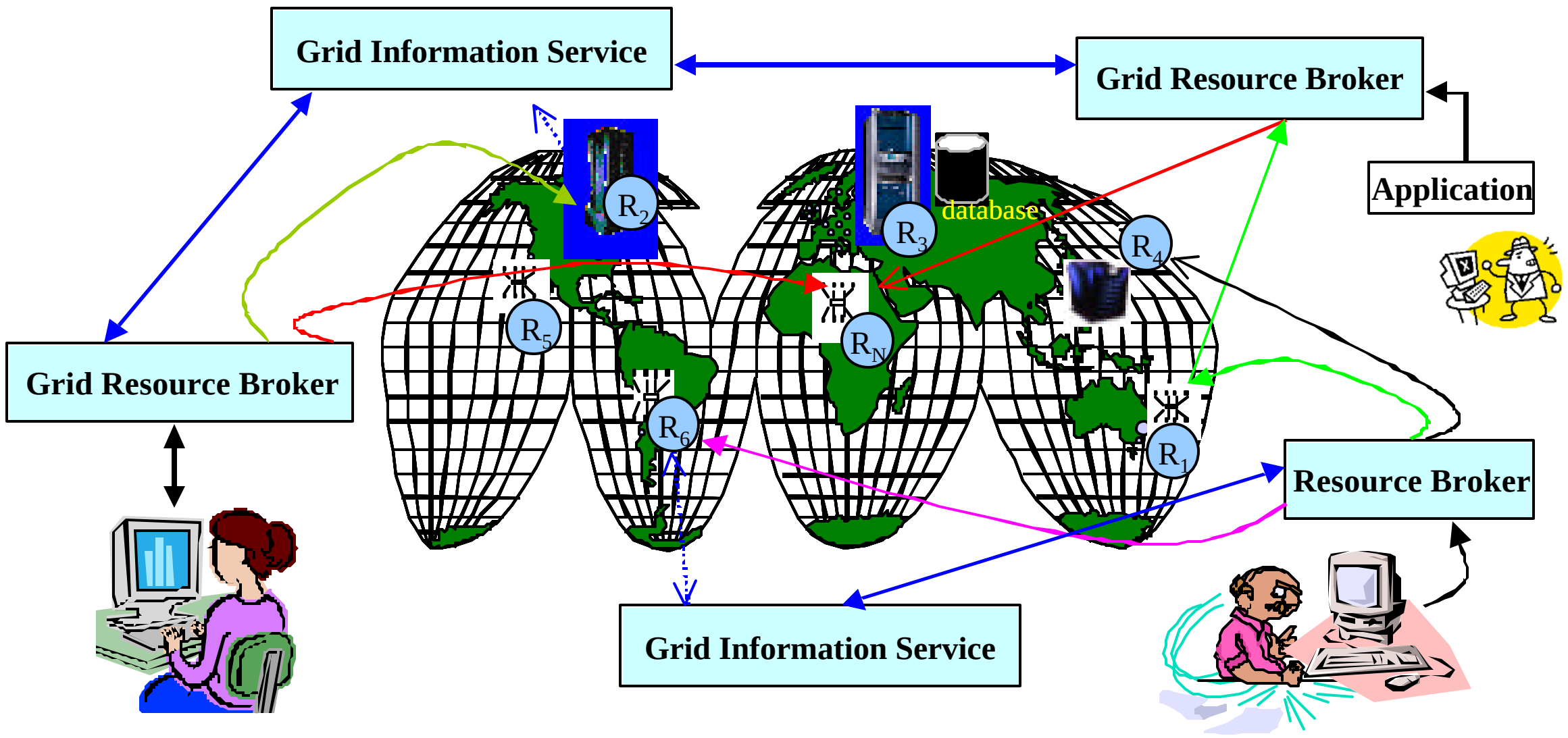

**Figure 1: A generic view of World-Wide Grid computing environment.**

The computing environments comprise heterogeneous resources (PCs, workstations, clusters, and supercomputers), fabric management systems (single system image OS, queuing systems, etc.) and policies, and applications (scientific, engineering, and commercial) with varied requirements (CPU, I/O, memory, and/or network intensive). The users: *producers* (also called *resource owners*) and *consumers* (also called *end-users*) have different goals, objectives, strategies, and demand patterns. More importantly both resources and end-users are geographically distributed with different time zones. In managing such complex grid environments, traditional approaches to resource management that attempt to optimize system-wide measures of performance cannot be employed. This is because traditional approaches use centralized policies that need complete state information and a common fabric management policy, or decentralized consensus based policy. In large-scale grid environments, it is impossible to define an acceptable system-wide performance matrix and common fabric management policy. Apart from the centralized approach, two other approaches that are used in distributed resource management are: *hierarchical* and *decentralized* scheduling or a combination of them [9]. We note that similar heterogeneity and decentralization complexities exist in human economies where market driven economic models have been used to successfully manage them. Therefore, in [9][10][11], we investigated on the use of economics as a metaphor for management of resources in grid computing environments.

A grid resource broker, called Nimrod-G [8], has been developed that performs scheduling of parameter sweep, task-farming applications on geographically distributed resources. It supports deadline and budget based scheduling driven by market-based economic models. To meet users' quality of service requirements, our broker dynamically leases grid resources and services at runtime depending on their capability, cost, and availability. Many scheduling experiments have been conducted on the execution of data-intensive, science applications such as molecular modeling for drug design under a few grid scenarios (like 2 hours deadline and 10 machines for a single user). The ability to experiment with a large number of grid scenarios was limited by the number of resources that were available in the WWG (World-Wide Grid) testbed [20]. Also, it was impossible to create a *repeatable* and *controlled* environment for experimentation and evaluation of scheduling strategies. This is because resources in the grid span across multiple administrative domains, each with their own policies, users, and priorities.

The researchers and students, investigating resource management and scheduling for large scale distributed computing, need a simple framework for deterministic modeling and simulation of resources and applications to evaluate scheduling strategies. For most who do not have access to ready-to-use testbed infrastructures, building them is expensive and time consuming. Also, even for those who have access, the testbed size is limited to a few resources and domains; and testing scheduling algorithms for scalability and adaptability, and evaluating scheduler performance for various applications and resource scenarios is harder and impossible to trace. To overcome these limitations, we provide a Java-based grid simulation toolkit called GridSim. The grid computing researchers and educators also recognized the importance and the need for such a toolkit for modeling and simulation environments [13]. It should be noted that this paper has a major orientation towards grid, however, we believe that our discussion and thoughts apply equally well to peer-to-peer (P2P) systems since resource management and scheduling issues in both systems are quite similar.

The GridSim toolkit supports modeling and simulation of a wide range of heterogeneous resources, such as single or multiprocessors, shared and distributed memory machines such as PCs, workstations, SMPs, and clusters with different capabilities and configurations. It can be used for modeling and simulation of application scheduling on various classes of parallel and distributed computing systems such as clusters [2], grids [3], and P2P networks [4]. The resources in clusters are located in a single administrative domain and managed by a single entity whereas, in grid and P2P systems, resources are geographically distributed across multiple administrative domains with their own management policies and goals. Another key difference between cluster and grid/P2P systems arises from the way application scheduling is performed. The *schedulers* in cluster systems focus on enhancing overall system performance and utility, as they are responsible for the whole system. Whereas, schedulers in grid/P2P systems called *resource brokers*, focus on enhancing performance of a specific application in such a way that its end-users requirements are met.

The GridSim toolkit provides facilities for the modeling and simulation of resources and network connectivity with different capabilities, configurations, and domains. It supports primitives for application composition, information services for resource discovery, and interfaces for assigning application tasks to resources and managing their execution. These features can be used to simulate resource brokers or Grid schedulers for evaluating performance of scheduling algorithms or heuristics. We have used GridSim toolkit to create a resource broker that simulates Nimrod-G for design and evaluation of deadline and budget constrained scheduling algorithms with cost and time optimizations.

The rest of this paper is organized as follows. Section 2 discusses related work with highlights on unique features that distinguish our toolkit from other packages. The GridSim architecture and internal components that make up GridSim simulations are discussed in Section 3. Section 4, discusses how to build GridSim based scheduling simulations. Sample results of simulation of a resource broker similar to Nimrod-G with deadline and budget constrained cost-optimization scheduling algorithm is discussed in Section 0. The final section summarizes the paper along with suggestions for future works.

## 2 Related Work

Simulation has been used extensively for modeling and evaluation of real world systems, from business process and factory assembly line to computer systems design. Accordingly over the years, modeling and simulation has emerged as an important discipline and many standard and application-specific tools and technologies have been built. They include simulation languages (e.g., Simscript [14]), simulation environments (e.g., Parsec [15]), simulation libraries (SimJava [1]), and application specific simulators (e.g., OMNet++ network simulator [19]). While there exists a large body of knowledge and tools, there are very few tools available for application scheduling simulation in Grid computing environments. The notable ones are: Bricks [16], MicroGrid [18], Simgrid [17], and our GridSim toolkit.

The Bricks simulation system [16], developed at the Tokyo Institute of Technology in Japan, helps in simulating client-server like global computing systems that provide remote access to scientific libraries and packages running on high performance computers. It follows centralized global scheduling methodology as opposed to our work in which each application scheduling is managed by the users' own resource broker.

The MicroGrid emulator [18], undertaken in the University of California at San Diego (UCSD), is modeled after Globus [12]. It allows execution of applications constructed using Globus toolkit in a controlled virtual grid emulated environment. The results produced by emulation can be precise, but modeling numerous applications, grid environments, and scheduling scenarios for realistic statistical analysis of scheduling algorithms is time consuming as applications run on emulated resources. Also, scheduling algorithms designers generally work with application models instead of constructing actual applications. Therefore, MicroGrid's need for an application constructed using Globus imposes significant development overhead. However, when an actual system is implemented by incorporating scheduling strategies that are evaluated using simulation, the MicroGrid emulator can be used as a complementary tool for verifying simulation results with real applications.

The Simgrid toolkit [17], developed in the University of California at San Diego (UCSD), is a C language based toolkit for the simulation of application scheduling. It supports modeling of resources that are *time-shared* and the load can be injected as constants or from real traces. It is a powerful system that allows creation of tasks in terms of their execution time and resources with respect to a standard machine capability. Using Simgrid APIs, tasks can be assigned to resources depending on the scheduling policy being simulated. It has been used for a number of real studies, and demonstrates the power of simulation. However, because Simgrid is restricted to a single scheduling entity and time-shared systems, it is difficult to simulate multiple competing users, applications, and schedulers, each with their own policies when operating under market like grid computing environment, without extending the toolkit substantially. Also, many large-scale resources in the grid environment are space-shared machines and they need to be supported in simulation. Hence, our GridSim toolkit extends the ideas in existing systems and overcomes their limitations accordingly.

Finally, we have chosen to implement GridSim in Java by leveraging SimJava's [1] basic discrete event simulation infrastructure. This feature is likely to appeal to educators and students since Java has emerged as a popular programming language for network computing.

## 3 GridSim: Grid Modeling and Simulation Toolkit

The GridSim toolkit provides a comprehensive facility for simulation of different classes of heterogeneous resources, users, applications, resource brokers, and schedulers. It can be used to simulate application schedulers for single or multiple administrative domains distributed computing systems such as clusters and grids. Application schedulers in grid environment, called resource brokers, perform resource discovery, selection, and aggregation of a diverse set of distributed resources for an individual user. That means, each user has his own private resource broker and hence, it can be targeted to optimize for the requirements and objectives of its owner. Whereas schedulers, managing resources such as clusters in a single administrative domain, have complete control over the policy used for allocation of resources. That means, all users need to submit their jobs to the *central* scheduler, which can be targeted to perform global optimization such as higher system utilization and overall user satisfaction depending on resource allocation policy or optimize for high priority users.

### 3.1 Features

Salient features of the GridSim toolkit include the following:

- It allows modeling of heterogeneous types of resources.
- Resources can be modeled operating under space- or time-shared mode.
- Resource capability can be defined (in the form of MIPS as per SPEC benchmark).
- Resources can be located in any time zone.
- Weekends and holidays can be mapped depending on resource's local time to model non-Grid (local) workload.
- Resources can be booked for advance reservation.
- Applications with different parallel application models can be simulated.
- Application tasks can be heterogeneous and they can be CPU or I/O intensive.
- There is no limit on the number of application jobs that can be submitted to a resource.

- Multiple user entities can submit tasks for execution simultaneously in the same resource, which may be time-shared or space-shared. This feature helps in building schedulers that can use different market-driven economic models for selecting services competitively.
- Network speed between resources can be specified.
- It supports simulation of both static and dynamic schedulers.
- Statistics of all or selected operations can be recorded and they can be analyzed using GridSim statistics analysis methods.

## 3.2 System Architecture

We employed a layered and modular architecture for grid simulation to leverage existing technologies and manage them as separate components. A multi-layer architecture and abstraction for the development of GridSim platform and its applications is shown in Figure 2. The first layer is concerned with the scalable Java's interface and the runtime machinery, called JVM (Java Virtual Machine), whose implementation is available for single and multiprocessor systems including clusters [25]. The second layer is concerned with a basic discrete-event infrastructure built using the interfaces provided by the first layer. One of the popular discrete-event infrastructure implementations available in Java is SimJava [1]. Recently a distributed implementation of SimJava is also made available. The third layer is concerned with modeling and simulation of core Grid entities such as resources, information services, and so on; application model, uniform access interface, and primitives application modeling and framework for creating higher level entities. The GridSim toolkit focuses on this layer that simulates system entities using the discrete-event services offered by the lower-level infrastructure. The fourth layer is concerned with the simulation of resource aggregators called grid resource brokers or schedulers. The final layer focuses on application and resource modeling with different scenarios using the services provided by the two lower-level layers for evaluating scheduling and resource management policies, heuristics, and algorithms. In this section, we briefly discuss SimJava model for discrete events (a second-layer component) and focus mainly on the GridSim (the third-layer) design and implementation. The resource broker simulation and performance evaluation is highlighted in the next two sections.

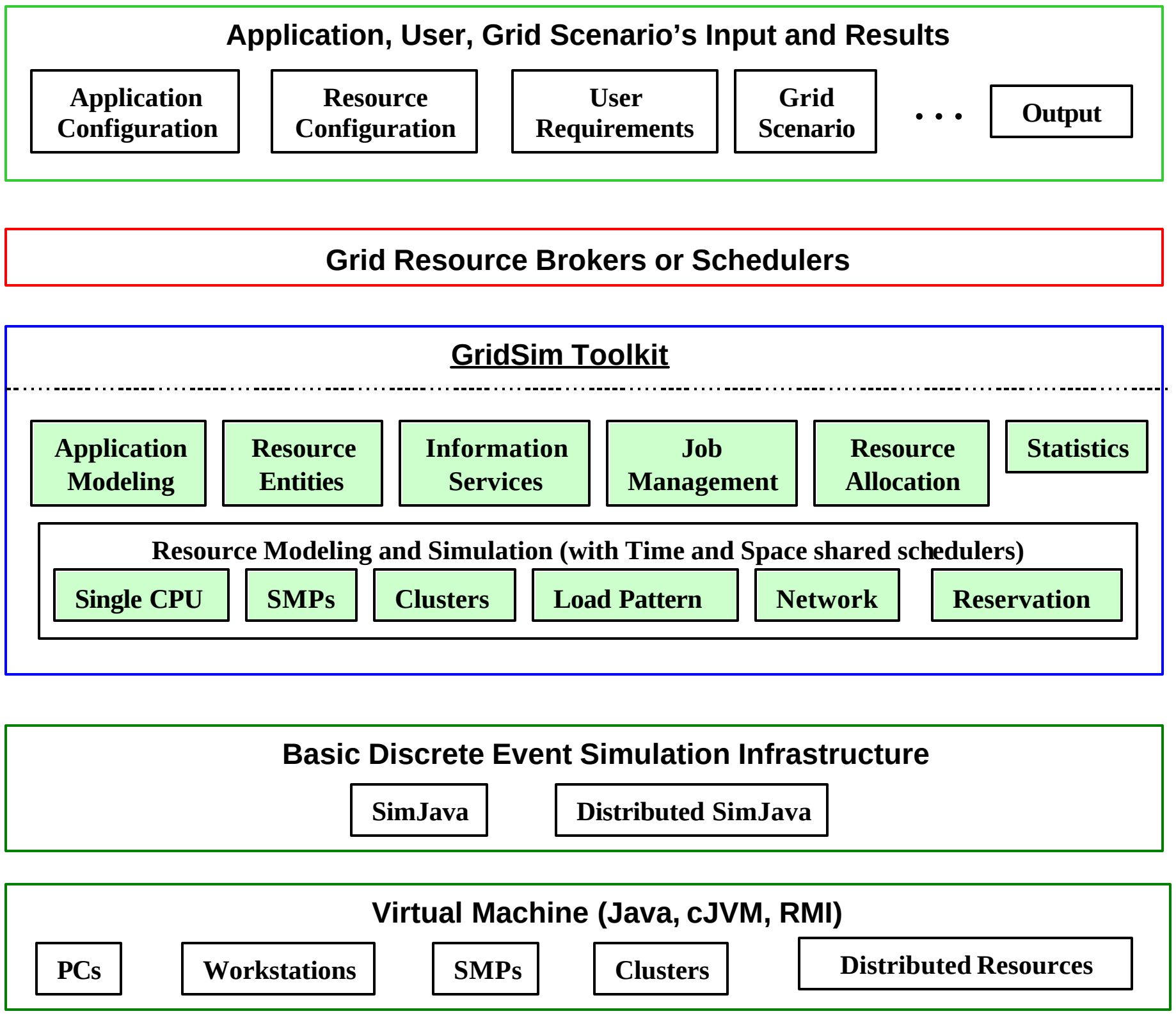

**Figure 2: A modular architecture for GridSim platform and components.**

### 3.2.1 SimJava Discrete Event Model

SimJava [1] is a general-purpose discrete event simulation package implemented in Java. Simulations in SimJava contain a number of entities each of which runs in parallel in its own thread. An entity's behaviour is encoded in Java using its body() method. Entities have access to a small number of simulation primitives:

- sim_schedule() sends event objects to other entities via ports;
- sim_hold() holds for some simulation time;
- sim_wait() waits for an event object to arrive.

These features help in constructing a network of active entities that communicate by sending and receiving passive event objects efficiently.

The sequential discrete event simulation algorithm, in SimJava, is as follows. A central object Sim_system maintains a timestamp ordered queue of future events. Initially all entities are created and their body() methods are put in run state. When an entity calls a simulation function, the Sim_system object halts that entity's thread and places an event on the future queue to signify processing the function. When all entities have halted, Sim_system pops the next event off the queue, advances the simulation time accordingly, and restarts entities as appropriate. This continues until no more events are generated. If the Java virtual machine supports native threads, then all entities starting at exactly the same simulation time may run concurrently.

### 3.2.2 GridSim Entities

GridSim supports entities for simulation of single processor and multiprocessor, heterogeneous resources that can be configured as time or space shared systems. It allows setting their clock to different time zones to simulate geographic distribution of resources. It supports entities that simulate networks used for communication among resources. During simulation, GridSim creates a number of multi-threaded entities, each of which runs in parallel in its own thread. An entity's behavior needs to be simulated within its body() method, as dictated by SimJava.

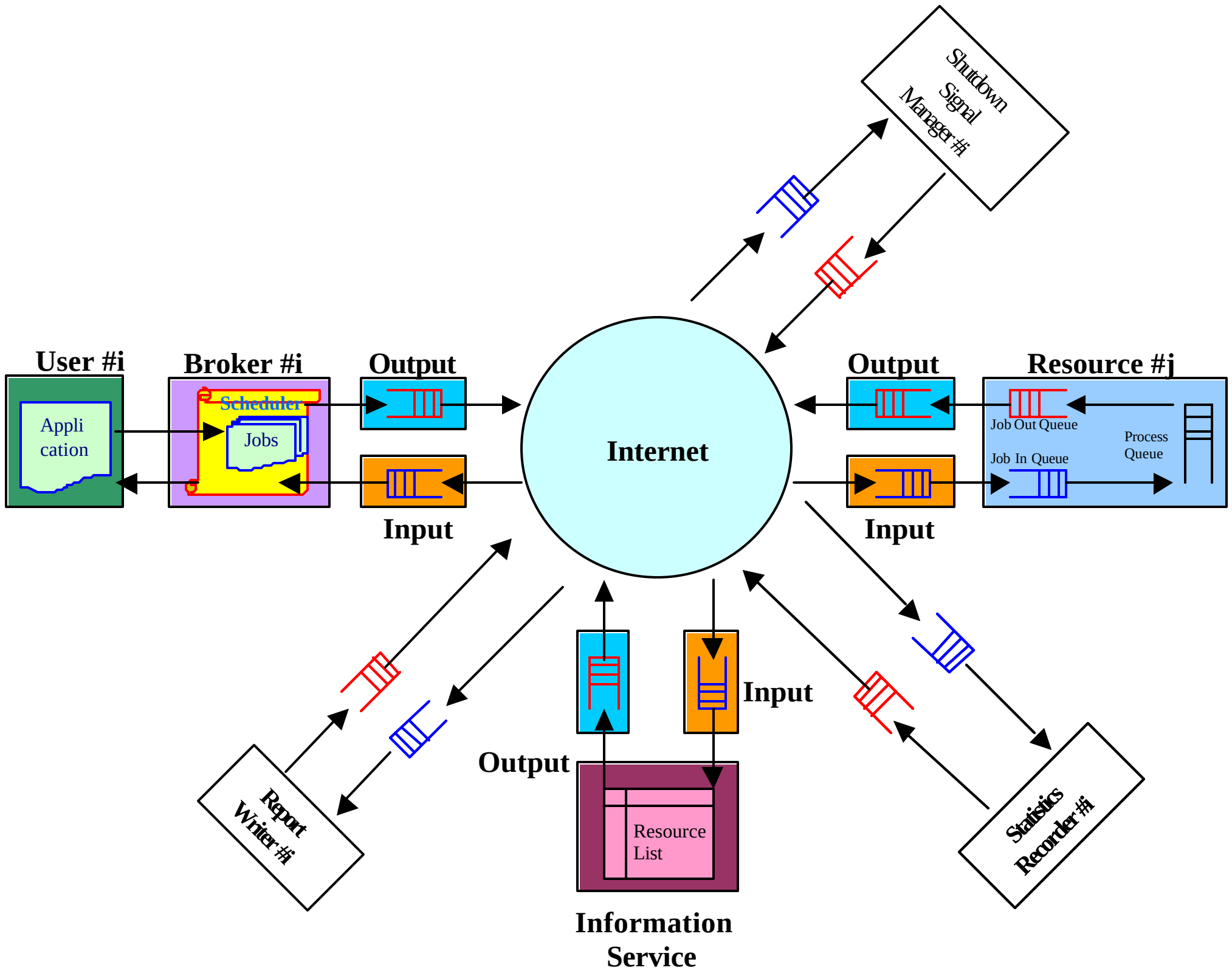

**Figure 3: A flow diagram in GridSim based simulations.**

A simulation environment needs to abstract all the entities and their time dependent interactions in the real system. It needs to support the creation of user-defined time dependent response functions for the interacting entities. The response function can be a function of the past, current, or both states of entities. GridSim based simulations contain entities for the users, brokers, resources, information service, statistics, and network based I/O as shown in Figure 3. The design and implementation issues of these GridSim entities are discussed below:

**User –** Each instance of the User entity represents a Grid user. Each user may differ from the rest of users with respect to the following characteristics:

- Types of job created e.g., job execution time, number of parametric replications, etc.,
- Scheduling optimization strategy e.g., minimization of cost, time, or both,
- Activity rate e.g., how often it creates new job,
- Time zone, and
- Absolute deadline and budget, or
- D-and B-factors, deadline and budget relaxation parameters, measured in the range [0,1] express deadline and budget affordability of the user relative to the application processing requirements and available resources.

**Broker –** Each user is connected to an instance of the Broker entity. Every job of a user is first submitted to its broker and the broker then schedules the parametric tasks according to the user's scheduling policy. Before scheduling the tasks, the broker dynamically gets a list of available resources from the global directory entity. Every broker tries to optimize the policy of its user and therefore, brokers are expected to face extreme competition while gaining access to resources. The scheduling algorithms used by the brokers must be highly adaptable to the market's supply and demand situation.

**Resource –** Each instance of the Resource entity represents a grid resource. Each resource may differ from the rest of resources with respect to the following characteristics:

- Number of processors;
- Cost of processing;
- Speed of processing;
- Internal process scheduling policy e.g., time shared or space shared;
- Local load factor; and
- Time zone.

The resource speed and the job execution time can be defined in terms of the ratings of standard benchmarks such as MIPS and SPEC. They can also be defined with respect to the standard machine. Upon obtaining the resource contact details from the grid information service, brokers can query resources directly for their static and dynamic properties.

**Grid Information Service –** It provides resource registration services and keeps track of a list of resources available in the Grid. The brokers can query this for resource contact, configuration, and status information.

**Input and Output –**The flow of information among the GridSim entities happen via their Input and Output entities. Every networked GridSim entity has I/O channels or ports, which are used for establishing a link between the entity and its own Input and Output entities. Note that the GridSim entity and its Input and Output entities are threaded entities i.e., they have their own execution thread with body() method that handles events. The architecture for entity communication model in GridSim is illustrated in Figure 4. The use of separate entities for input and output enables a networked entity to model full duplex and multi-user parallel communications. The support for buffered input and output channels associated with every GridSim entity provides a simple mechanism for an entity to communicate with other entities and at the same time enables modeling of necessary communications delay transparently.

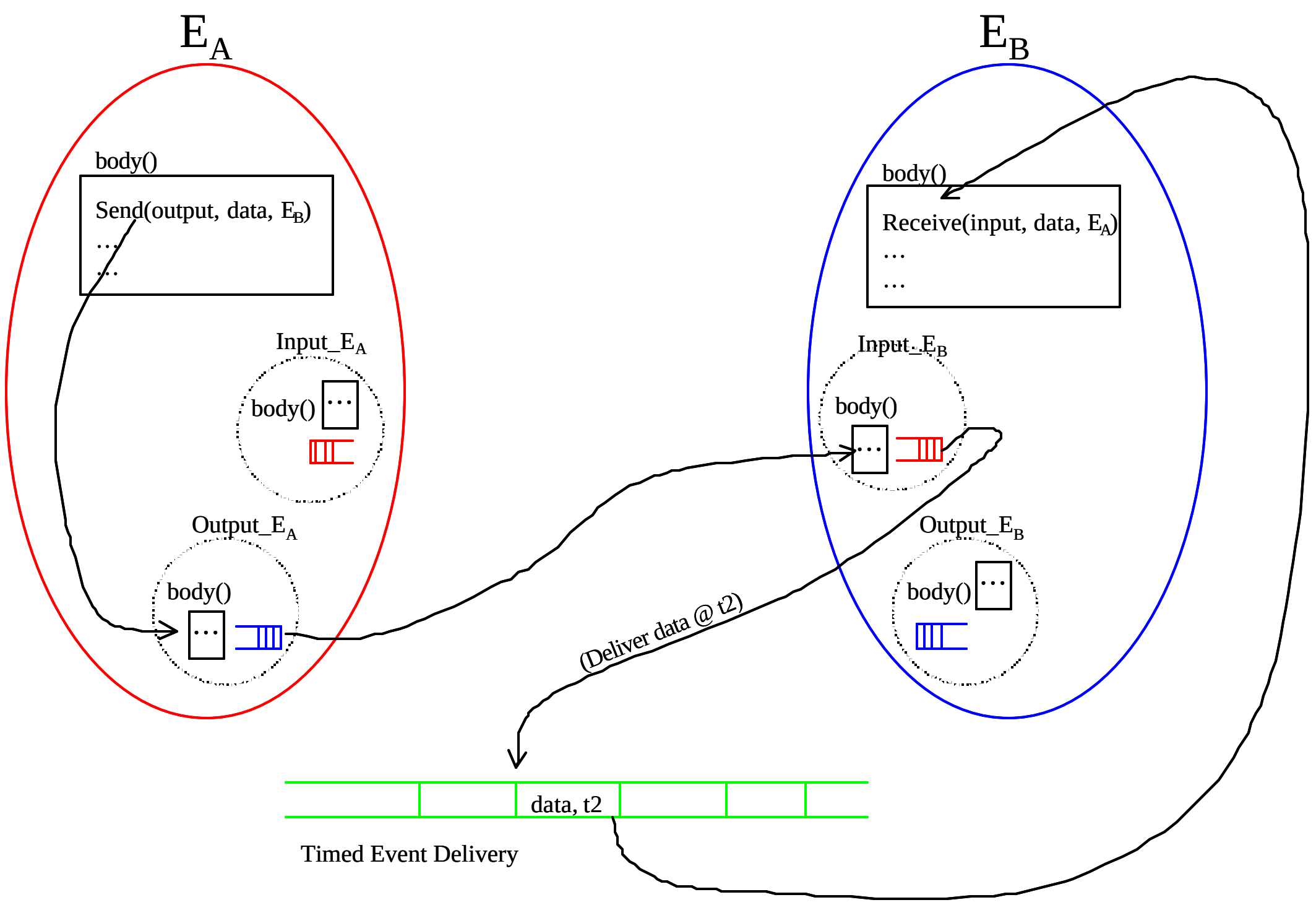

**Figure 4: Entity communication model via its Input and Output entities.**

## 3.3 Application Model

GridSim does not explicitly define any specific application model. It is up to the developers (of schedulers and resource brokers) to define them. We have experimented with a task-farming application model and we believe that other parallel application models such as process parallelism, DAGs (Directed Acyclic Graphs), divide and conquer etc., described in [5], can also be modeled and simulated using GridSim.

In GridSim, each independent task may require varying processing time and input files size. Such tasks can be created and their requirements are defined through *Gridlet* objects. A *Gridlet* is a package that contains all the information related to the job and its execution management details such as job length expressed in MIPS, disk I/O operations, the size of input and output files, and the job originator. These basic parameters help in determining execution time, the time required to transport input and output files between users and remote resources, and returning the processed Gridlets back to the originator along with the results. The GridSim toolkit supports a wide range of Gridlet management protocols and services that allow schedulers to map a Gridlet to a resource and manage it through out the life cycle.

## 3.4 Interaction Protocols Model

The protocols for interaction between GridSim entities are implemented using events. In GridSim, entities use events for both service request and service delivery. The events can be raised by any entity to be delivered immediately or with specified delay to other entities or itself. The events that are originated from the same entity are called *internal events* and those originated from the external entities are called *external events*. Entities can distinguish these events based on the source identification associated with them. The GridSim protocols are used for defining entity services. Depending on the service protocols, the GridSim events can be further classified into *synchronous* and *asynchronous* events. An event is called *synchronous* when the event source entity waits until the event destination entity performs all the actions associated with the event (i.e., the delivery of full service). An event is called *asynchronous* when the event source entity raises an event and continues with other activities without waiting for its completion. When the destination entity receives such events or service requests, it responds back with results by sending one or more events, which can then take appropriate actions. It should be noted that external events could be synchronous or asynchronous, but internal events need to be raised as asynchronous events only to avoid deadlocks.

A complete set of entities in a typical GridSim simulation and the use of events for simulating interaction between them are shown in Figure 5 and Figure 6. Figure 5 emphasizes the interaction between a resource entity that simulates time-shared scheduling and other entities. Figure 6 emphasizes the interaction between a resource entity that simulates space-shared system and other entities. In this section we briefly discuss the use of events for simulating grid activities.

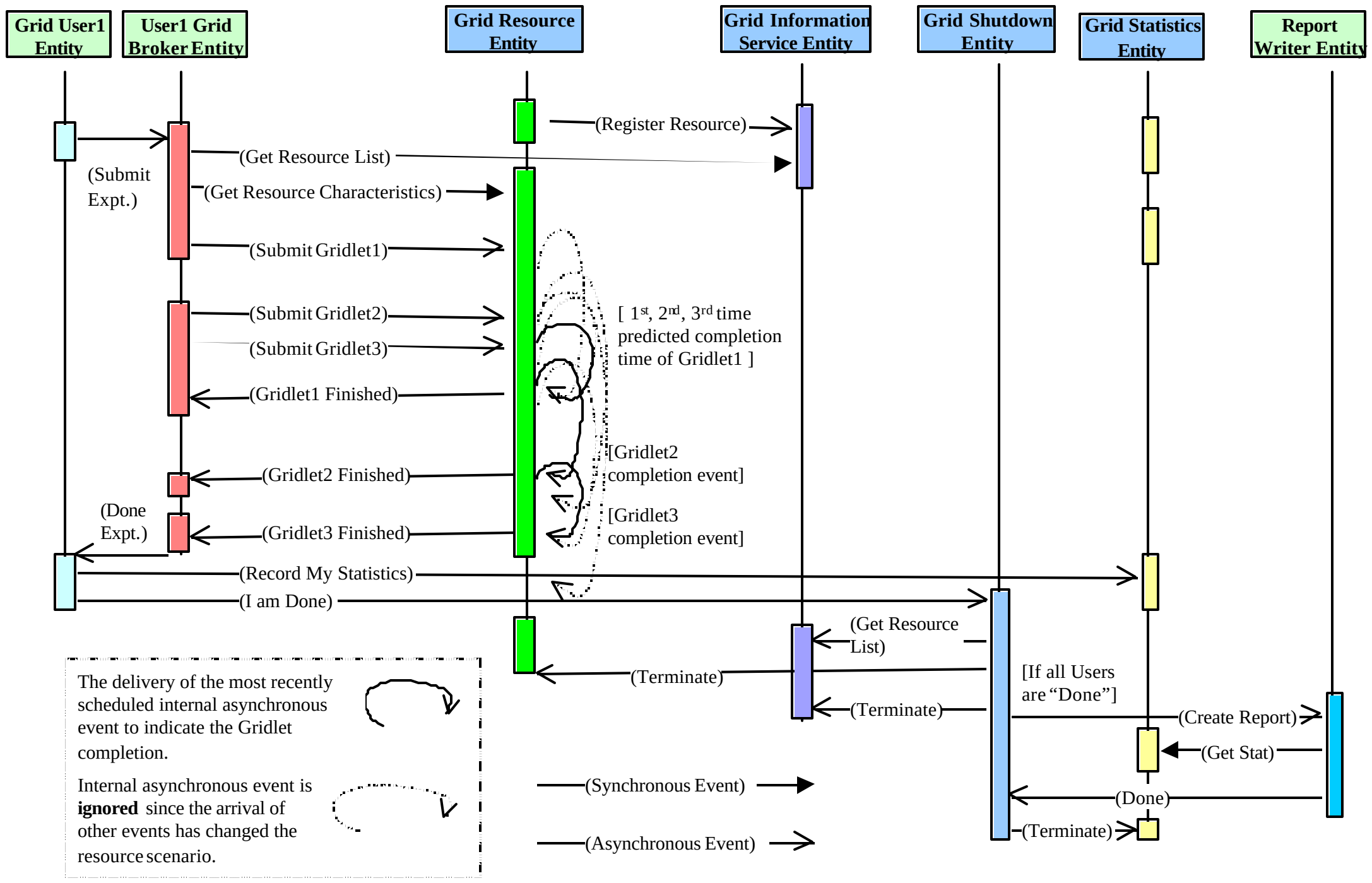

**Figure 5: An event diagram for interaction between a time-shared resource and other entities.**

The GridSim entities (user, broker, resource, information service, statistics, shutdown, and report writer) send events to other entities to signify the request for service, deliver results, or raise internal actions. Note that GridSim implements core entities that simulate resource, information service, statistics, and shutdown services. These services are used to simulate user with application, broker for scheduling, and an optional report writer for creating statistical reports at the end of a simulation. The event source and destination entities must agree upon the protocols for service request and delivery. The protocols for interaction between the user-defined and core entities are pre-defined.

When GridSim starts, the resource entities register themselves with the Grid Information Service (GIS) entity, by sending events. This resource registration process is similar to GRIS (Grid Resource Information Server) registering with GIIS (Grid Index Information Server) in Globus system. Depending on the user entity's request, the broker entity sends an event to the GIS entity, to signify a query for resource discovery. The GIS entity returns a list of registered resources and their contact details. The broker entity sends events to resources with request for resource configuration and properties. They respond with dynamic information such as resources cost, capability, availability, load, and other configuration parameters. These events involving the GIS entity are synchronous in nature.

Depending on the resource selection and scheduling strategy, the broker entity places asynchronous events for resource entities in order to dispatch Gridlets for execution—the broker need not wait for a resource to complete the assigned work. When the Gridlet processing is finished, the resource entity updates the Gridlet status and processing time and sends it back to the broker by raising an event to signify its completion.

The GridSim resources use internal events to simulate resource behavior and resource allocation. The entity needs to be modeled in such a way that it is able to receive all events meant for it. However, it is up to the entity to decide on the associated actions. For example, in time-shared resource simulations (see Figure 5) internal events are scheduled to signify the completion time of a Gridlet, which has the smallest remaining processing time requirement. Meanwhile, if an external event arrives, it changes the share resource availability for each Gridlet. That means the most recently scheduled event may not necessarily signify the completion of a Gridlet. The resource entity can discard such internal events without processing. The use of internal events for simulating resources is discussed in detail in Section 3.5.

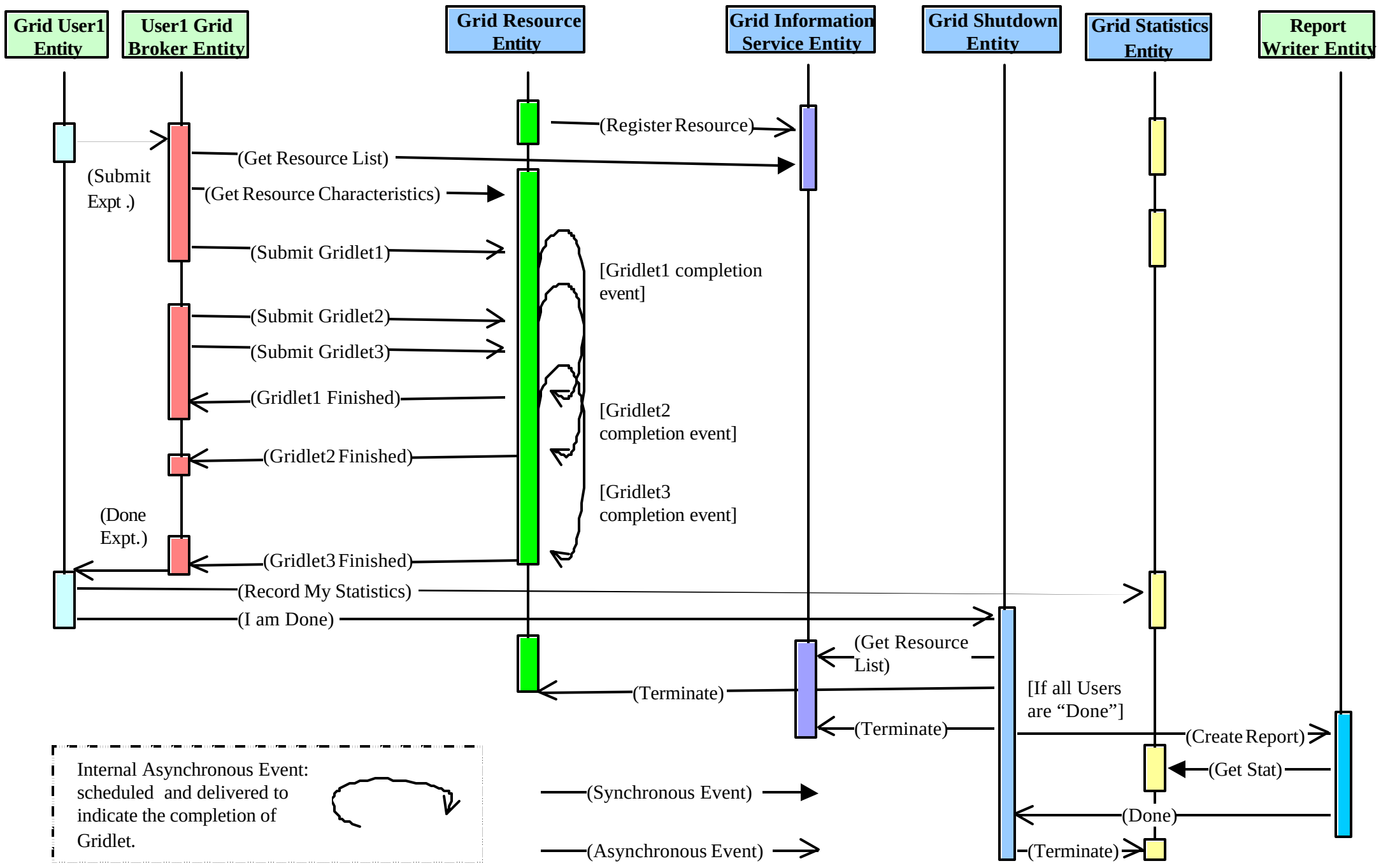

**Figure 6: An event diagram for interaction between a space-shared resource and other entities.**

## 3.5 Resource Model – Simulating Multitasking and Multiprocessing

In the GridSim toolkit, we can create Processing Elements (PEs) with different speeds (measured in either MIPS or SPEC-like ratings). Then, one or more PEs can be put together to create a machine. Similarly, one or more machines can be put together to create a Grid resource. Thus, the resulting Grid resource can be a single processor, shared memory multiprocessors (SMP), or a distributed memory cluster of computers. These grid resources can simulate time- or space-shared scheduling depending on the allocation policy. A single PE or SMP type grid resource is typically managed by time-shared operating systems that use round-robin scheduling policy (see Figure 9) for multitasking. The distributed memory multiprocessing systems (such as clusters) are managed by queuing systems, called space-shared schedulers, that execute a Gridlet by running it on a dedicated PE (see Figure 12) when allocated. The space-shared systems use resource allocation policies such as first-come-first-served (FCFS), back filling, shortest-job-first served (SJFS), and so on. It should also be noted that resource allocation within high-end SMPs could also be performed using the space-shared schedulers.

Multitasking and multiprocessing systems allow concurrently running tasks to share system resources such as processors, memory, storage, I/O, and network by scheduling their use for very short time intervals. A detailed simulation of scheduling tasks in the real systems would be complex and time consuming. Hence, in GridSim, we abstract these physical entities and simulate their behavior using process oriented, discrete event "interrupts" with time interval as large as the time required for the completion of a smallest remaining-time job. The GridSim resources can send, receive, or schedule events to simulate the execution of jobs. It schedules self-events for simulating resource allocation depending on the scheduling policy and the number of jobs in queue or in execution.

Let us consider the following scenario to illustrate the simulation of Gridlets execution and scheduling within a GridSim resource. A resource consists of two shared or distributed memory PEs each with MIPS rating of 1, for simplicity. Three Gridlets that represent jobs with processing requirements equivalent to 10, 8.5, and 9.5 MI (million instructions) arrive in simulation times 0, 4, and 7 respectively. The way GridSim schedules jobs to PEs is shown schematically in Figure 9 for time-shared resources and Figure 12 for space-shared resources.

### 3.5.1 Simulation of Scheduling in Time-Shared Resources

The GridSim resource simulator uses internal events to simulate the execution and allocation of PEs share to Gridlet jobs. When jobs arrive, time-shared systems start their execution immediately and share resources among all jobs. Whenever a new Gridlet job arrives, we update the processing time of existing Gridlets and then add this newly arrived job to the execution set. We schedule an internal event to be delivered at the earliest completion time of smallest job in the execution set. It then waits for the arrival of events.

A complete algorithm for simulation of time-share scheduling and execution is shown in Figure 7. If a newly arrived event happens to be an internal event whose tag number is the same as the most recently scheduled event, then it is recognized as a job completion event. Depending on the number of Gridlets in execution and the number of PEs in a resource, GridSim allocates appropriate amount of PE share to all Gridlets for the event duration using the algorithm shown in Figure 8. It should be noted that Gridlets sharing the same PE would get an equal amount of PE share. The completed Gridlet is sent back to its originator (broker or user) and removed from the execution set. GridSim schedules a new internal event to be delivered at the forecasted earliest completion time of the remaining Gridlets.

*Algorithm: Time-Shared Grid Resource Event Handler ()*

1. Wait for an event
2. If the external and Gridlet arrival event, then:
   BEGIN /* a new job has arrived */
   a. Allocate PE Share for Gridlets Processed so far
   b. Add arrived Gridlet to Execution_Set
   c. Forecast completion time of all Gridlets in Execution_Set
   d. Schedule an event to be delivered at the smallest completion time
   END
3. If event is internal and its tag value is the same as the recently scheduled internal event tag,
   BEGIN /* a job finish event */
   a. Allocate PE Share of all Gridlets processed so far
   b. Update finished Gridlet's PE and Wall clock time parameters and send it back to the broker
   c. Remove finished Gridlet from the Execution_Set and add to Finished_Set
   d. Forecast completion time of all Gridlets in Execution_Set
   e. Schedule an event to be delivered at the smallest completion time
   END
4. Repeat the above steps until the end of simulation event is received

**Figure 7: An event handler for simulating time-shared resource scheduling.**

Figure 9 illustrates the simulation of time-share scheduling algorithm and the Gridlets' execution. When Gridlet1 arrives at time 0, it is mapped to PE1 and an internal event to be delivered at the time 10 is scheduled since the predicted completion time is still 10. At time 4, Gridlet2 arrives and it is mapped to the PE2. The completion time of Gridlet2 was predicted as 12.5 and the completion time of Gridlet1 is still 10 since both of them are executing on different PEs. A new internal event is scheduled, which will still be delivered at time 10. At time 7, Gridlet3 arrives, which is mapped to the PE2. It shares the PE time with Gridlet2. At time 10, an internal event is delivered to the resource to signify the completion of the Gridlet1, which is then sent back to the broker. At this moment, as the number of Gridlets equal the number of PEs, they are mapped to different PEs. An internal event to be delivered at time 14 is scheduled to indicate the predicted completion time of Gridlet2. As simulation proceeds, an internal event is delivered at time 14 and Gridlet2 is sent back to the broker. An internal event to be delivered at time 18 is scheduled to indicate the predicted completion time of Gridlet3. Since there were no other Gridlets submitted before this time, the resource receives an internal interrupt at time 18, which signifies the completion of Gridlet3. A schematic representation of Gridlets arrival, internal events delivery, and sending them back to the broker is shown in Figure 5. A detailed statistical data on the arrival, execution start, finish, and elapsed time of all Gridlets is shown in Table 1.

*Algorithm: PE_Share_Allocation(Duration)*

1. Identify total MI per PE for the duration and the number of PE that process one extra Gridlet
   TotalMIperPE = MIPSRatingOfOnePE()*Duration
   MinNoOfGridletsPerPE = NoOfGridletsInExec / NoOfPEs
   NoOfPEsRunningOneExtraGridlet = NoOfGridletsInExec % NoOfPEs
2. Identify maximum and minimum MI share that Gridlet get in the Duration
   If(NoOfGridletsInExec <= NoOfPEs), then:
   MaxSharePerGridlet = MinSharePerGridlet = TotalMIperPE
   MaxShareNoOfGridlets = NoOfGridletsInExec
   Else /* NoOfGridletsInExec > NoOfPEs */
   MaxSharePerGridlet = TotalMIperPE/ MinNoOfGridletsPerPE
   MinSharePerGridlet = TotalMIperPE/(MinNoOfGridletsPerPE+1)
   MaxShareNoOfGridlets = (NoOfPEs - NoOfPEsRunningOneExtraGridlet)* MinNoOfGridletsPerPE

**Figure 8: PE share allocation to Gridlet in time-shared GridSim resource.**

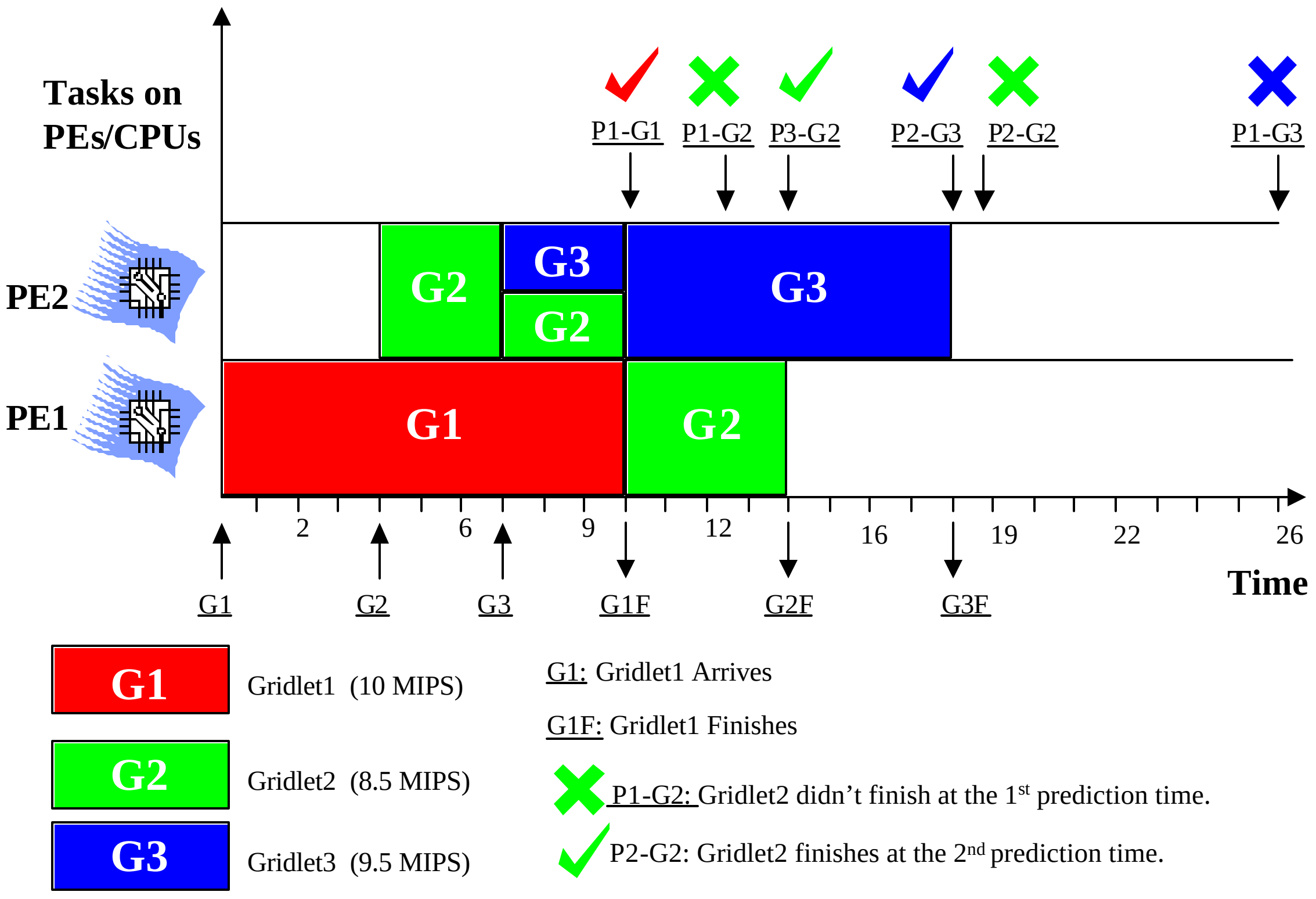

**Figure 9: Modeling time-shared multitasking and multiprocessing based on an event scheme.**

**Table 1: A scheduling statistics scenario for time- and space-shared resources in GridSim.**

| | | | Time-Shared Resource | | | Space-Shared Resource | | |
|---|---|---|---|---|---|---|---|---|
| **Grilets Number** | **Length (MI)** | **Arrival Time (a)** | **Start Time (s)** | **Finish Time (f)** | **Elapsed Time (f-a)** | **Start Time (s)** | **Finish Time (f)** | **Elapsed Time (f-a)** |
| G1 | 10 | 0 | 0 | 10 | 10 | 0 | 10 | 10 |
| G2 | 8.5 | 4 | 4 | 14 | 10 | 4 | 12.5 | 8.5 |
| G3 | 9.5 | 7 | 7 | 18 | 11 | 10 | 19.5 | 12.5 |

### 3.5.2 Simulation of Scheduling in Space-Shared Resources

The GridSim resource simulator uses internal events to simulate the execution and allocation of PEs to Gridlet jobs. When a job arrives, space-shared systems start its execution immediately if there is a free PE available, otherwise, it is queued. During the Gridlet assignment, job-processing time is determined and event is scheduled for delivery at the end of execution time. Whenever Gridlet job finishes and internal event is delivered to signify the completion of scheduled Gridlet job, the resource simulator frees the PE allocated to it and then checks if there are any other jobs waiting in the queue. If there are jobs waiting in the queue, then it selects a suitable job depending on the policy and assigns to the PE, which is free.

A complete algorithm for simulation of space-share scheduling and execution is shown in Figure 10. If newly arrived event happens to be an internal event whose tag number is the same as the most recently scheduled event, then it is recognized as a Gridlet completion event. If there are Gridlets in the submission queue, then depending on the allocation policy (e.g., the first Gridlet in the queue if FCFS policy is used), GridSim selects suitable Gridlets from the queue and assigns it to the PE or a suitable PE if more than one PE is free. See Figure 12 for illustration of the allocation of PE to Gridlets. The completed Gridlet is sent back to its originator (broker or user) and removed from the execution set. GridSim schedules a new internal event to be delivered at the completion time of the scheduled Gridlet job.

*Algorithm: Space-Shared Grid Resource Event Handler ()*

1. Wait for event and Identity Type of Event received
2. If it external and Gridlet arrival event, then:
   BEGIN /* a new job arrived */
   - If the number of Gridlets in execution are less than the number of PEs in the resource, then
     Allocate_PE_to_the_Gridlet() /* It should schedule an Gridlet completion event */
   - If not, Add Gridlet to the Gridlet_Submitted_Queue

   END
3. If event is internal and its tag value is the same recently scheduled internal event tag,
   BEGIN /* a job finish event */
   - Update finished Gridlet's PE and Wall clock time parameters and send it back to the broker
   - Set the status of PE to FREE
   - Remove finished Gridlet from the Execution_Set and add to Finished_Set
   - If Gridlet_Submitted_Queue has Gridlets in waiting, then
     Choose the Gridlet to be Processed() /* e.g., first one in Q if FCFS policy is used */
     Allocate_PE_to_the_Gridlet() /* It should schedule an Gridlet completion event */

   END
4. Repeat the above steps until the end of simulation event is received

**Figure 10: An event handler for simulating space-shared resource scheduling.**

*Algorithm: Allocate_PE_to_the_Gridlet(Gridlet gl)*

1. Identify a suitable Machine with Free PE
2. Identify a suitable PE in the machine and Assign to the Gridlet
3. Set Status of the Allocated PE to BUSY
4. Determine the Completion Time of Gridlet and Set an internal event to be delivered at the completion time

**Figure 11: PE allocation to the Gridlets in space-shared GridSim resource.**

Figure 12 illustrates the simulation of space-share scheduling algorithm and Gridlets' execution. When Gridlet1 arrives at time 0, it is mapped to PE1 and an internal event to be delivered at the time 10 is scheduled since the predicted completion time is still 10. At time 4, Gridlet2 arrives and it is mapped to the PE2. The completion time of Gridlet2 is predicted as 12.5 and the completion time of Gridlet1 is still 10 since both of them are executing on different PEs. A new internal event to be delivered at time 12.5 is scheduled to signify the completion of Gridlet2. At time 7, Gridlet3 arrives. Since there is no free PE available on the resource, it is put into to the queue. The simulation continues i.e., GridSim resource waits for the arrival of a new event. At time 10 a new event is delivered which happens to signify the completion of Gridlet1, which is then sent back to the broker. It then checks to see if there are any Gridlets waiting in the queue and chooses a suitable Gridlet (in this case as Gridlet2 is based on FCFS policy) and assign the available PE to it. An internal event to be delivered at time 19.5 is scheduled to indicate the completion time of Gridlet3 and then waits for the arrival of new events. A new event is delivered at the simulation time 12.5, which signifies the completion of the Gridlet2, which is then sent back to the broker. There is no Gridlet waiting in the queue, so it proceeds without scheduling any events and waits for the arrival of the next event. A new internal event arrives at the simulation time 19.5, which signifies the completion of Gridlet3. This process continues until resources receive an external event indicating the termination of simulation. A schematic representation of Gridlets arrival, internal events delivery, and sending them back to the broker is shown in Figure 6. A detailed statistical data on the arrival, execution start, finish, and elapsed time of all Gridlets is shown in Table 1.

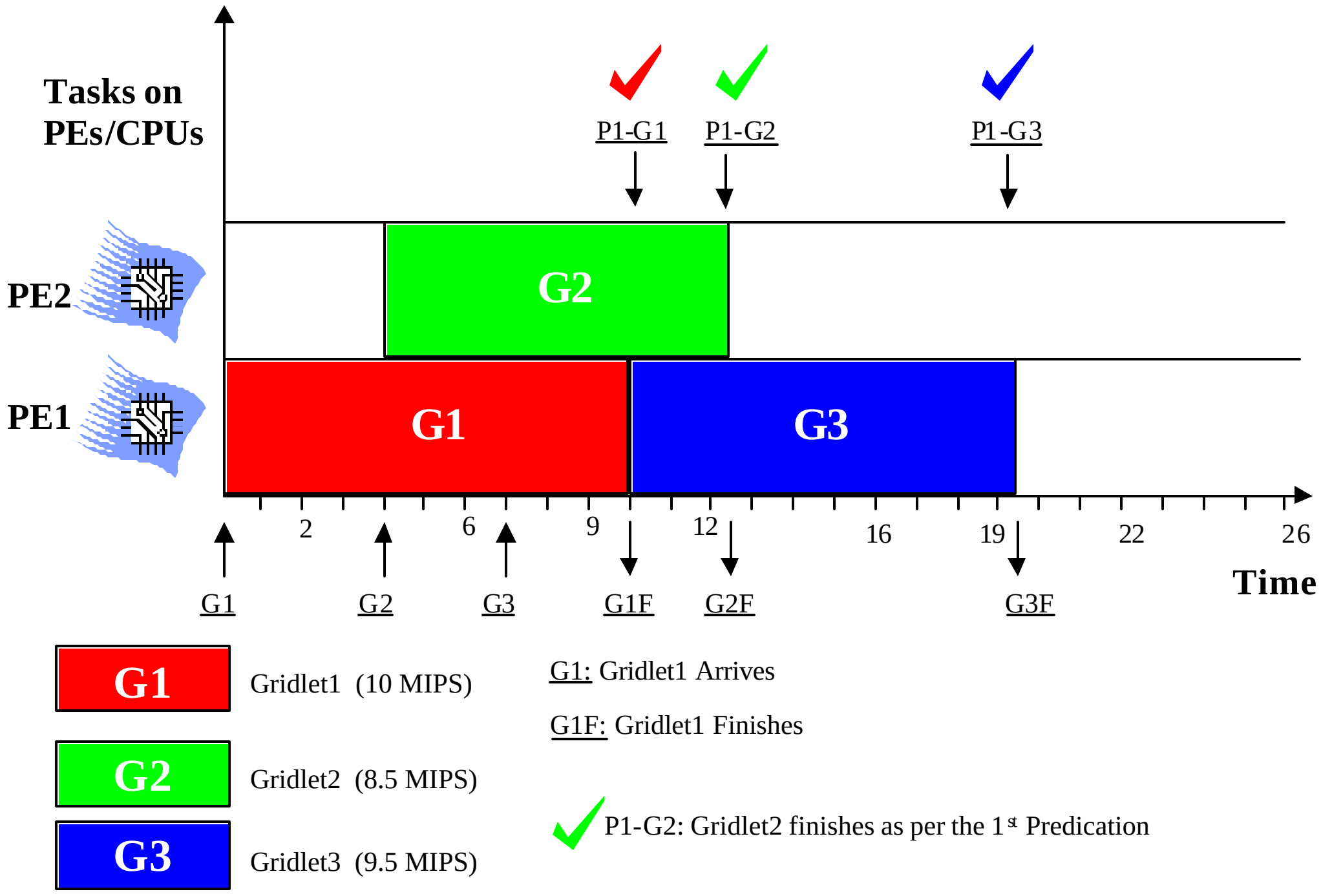

**Figure 12: Modeling space-shared multiprocessing based on an event scheme.**

For every Grid resource, the non-Grid (local) workload is estimated based on typically observed load conditions depending on the time zone of the resource. The network communication speed between a user and the resources is defined in terms of a data transfer speed (baud rate).

## 3.6 GridSim Java Package Design

A class diagram hierarchy of the gridsim package, represented using unified modeling language (UML) notation, is shown in Figure 13. The specification of each class contains up to three parts: attributes, methods, and internal classes. In the class diagram, attributes and methods are prefixed with characters "+", "-", and "#" indicating access modifiers public, private, and protected respectively. The gridsim package implements the following classes:

`class gridsim.Input` – This class extends the eduni.simjava.Sim_entity class. This class defines a port through which a simulation entity receives data from the simulated network. It maintains an event queue to serialize the data-in-flow and delivers to its parent entity. Simultaneous inputs can be modeled using multiple instances of this class.

`class gridsim.Output` – This class is very similar to the gridsim.Input class and it defines a port through which a simulation entity sends data to the simulated network. It maintains an event queue to serialize the data-out-flow and delivers to the destination entity. Simultaneous outputs can be modeled by using multiple instances of this class.

`class gridsim.GridSim` – This is the main class of Gridsim package that must be extended by GridSim entities. It inherits event management and threaded entity features from the eduni.simjava.Sim_entity class. The GridSim class adds networking and event delivery features, which allows synchronous or asynchronous communication for service access or delivery. All classes that extend the GridSim class must implement a method called "body()", which is automatically invoked since it is expected to be responsible for simulating entity behavior. The entities that extend the GridSim class can be instantiated with or without networked I/O ports. A networked GridSim entity gains communication capability via the objects of GridSim's I/O entity classes gridsim.Input and gridsim.Output classes. Each I/O entity will have a unique name assuming each GridSim entity that the user creates has unique name. For example, a resource entity with name "Resource2" will have an input entity whose name is prefixed with "Input_", making input entity full name as "Input_Resource2", which is expected to be unique. The I/O entities are concurrent entities, but they are visible within GridSim entity and are able to communicate with other GridSim entities by sending messages.

The GridSim class supports methods for simulation initialization, management, and flow control. The GridSim environment must be initialized to setup simulation environment before creating any other GridSim entities at the user level. This method also prepares the system for simulation by creating three GridSim internal entities—GridInformationService, GridSimShutdown, and GridStatistics. As explained in Section 3.2, the GridInformationService entity simulates the directory that dynamically keeps a list of resources available in the Grid. The GridSimShutdown entity helps in wrapping up a simulation by systematically closing all the opened GridSim entities. The GridStatistics entity provides standard services during the simulation to accumulate statistical data. Invoking the GridSim.Start () method starts the Grid simulation. All the resource and user entities must be instantiated in between invoking the above two methods.

The GridSim class supports static methods for sending and receiving messages between entities directly or via network entities, managing and accessing handle to various GridSim core entities, and recording statistics.

`class gridsim.PE` – It is used to represent CPU/*Processing Element* (PE) whose capability is defined in terms of MIPS rating.

`class gridsim.PEList` – It maintains a list of PEs that make up a machine.

`class gridsim.Machine` – It represents a uniprocessor or shared memory multiprocessor machine.

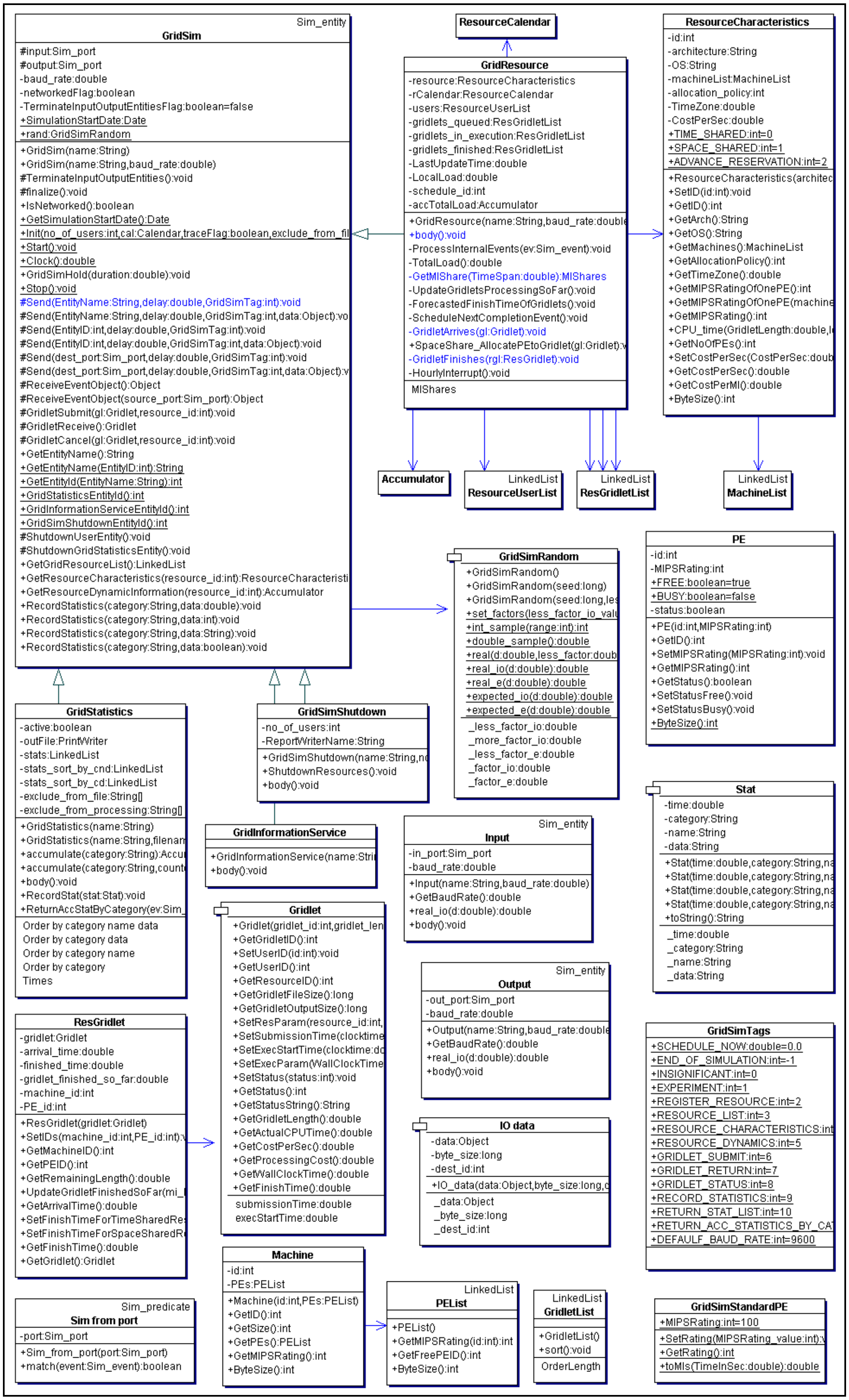

Figure 13: A class hierarchy diagram of GridSim package.

`class gridsim.MachineList` – An instance of this class simulates a collection of machines. It is up to the GridSim users to define the connectivity among the machines in a collection. Therefore, this class can be instantiated to model simple LAN to cluster to WAN.

`class gridsim.ResourceCharacteristics` –It represents static properties of a resource such as resource architecture, OS, management policy (time or space shared), cost, and time zone at which the resource is located along resource configuration.

`class gridsim.GridResource` – It extends the GridSim class and gains communication and concurrent entity capability. An instance of this class simulates a resource with properties defined in an object of the `gridsim.ResourceCharacteristics` class. The process of creating a Grid resource is as follows: first create PE objects with a suitable MIPS/SPEC rating, second assemble them together to create a machine. Finally, group one or more objects of the Machine to form a resource. A resource having a single machine with one or more PEs is managed as a time-shared system using round robin scheduling algorithm. A resource with multiple machines is treated as a distributed memory cluster and is managed as a space-shared system using first-come first served scheduling policy or its variants.

`class gridsim.GridSimStandardPE` – It defines MIPS rating for a standard PE or enables the users to define their own MIPS/SPEC rating for a standard PE. This value can be used for creating PEs with relative MIPS/SPEC rating for GridSim resources and creating Gridlets with relative processing requirements.

`class gridsim.ResourceCalendar` – This class implements a mechanism to support modeling local load on Grid resources that may vary according to the time zone, time, weekends, and holidays.

`class gridsim.GridInformationService` – A GridSim entity that provides grid resource registration, indexing and discovery services. The grid resources register their readiness to process Gridlets by registering themselves with this entity. GridSim entities such as the resource broker can contact this entity for resource discovery service, which returns a list of registered resource entities and their contact address. For example, scheduling entities use this service for resource discovery.

`class gridsim.Gridlet` – This class acts as job package that contains job length in MI, the length of input and out data in bytes, execution start and end time, and the originator of job. Individual users model their application by creating Gridlets for processing them on grid resources assigned by scheduling entities (resource brokers).

`class gridsim.GridletList` – It can be used to maintain a list of Gridlets and supports methods for organizing them.

`class gridsim.GridSimTags` – It contains various static command tags that indicate a type of action that needs to be undertaken by GridSim entities when they receive events. The different types of tags supported in GridSim along with comments indicating possible purpose are shown in Figure 14.

```
public class GridSimTags {
  public static final double SCHEDULE_NOW = 0.0;  // 0.0 indicates NO delay
  public static final int END_OF_SIMULATION = -1;
  public static final int INSIGNIFICANT = 0; // ignore tag
  public static final int EXPERIMENT = 1; // User <-> Broker
  public static final int REGISTER_RESOURCE = 2; // GIS -> ResourceEntity
  public static final int RESOURCE_LIST = 3; // GIS <-> Broker
  public static final int RESOURCE_CHARACTERISTICS = 4; // Broker <-> ResourceEntity
  public static final int RESOURCE_DYNAMICS = 5; // Broker <-> ResourceEntity
  public static final int GRIDLET_SUBMIT = 6; // Broker -> ResourceEntity
  public static final int GRIDLET_RETURN = 7; // Broker <- ResourceEntity
  public static final int GRIDLET_STATUS = 8; // Broker <-> ResourceEntity
  public static final int RECORD_STATISTICS = 9; // Entity -> GridStatistics
  public static final int RETURN_STAT_LIST = 10; // Entity <- GridStatistics
  public static final int RETURN_ACC_STATISTICS_BY_CATEGORY = 11;
  public static final int DEFAULF_BAUD_RATE = 9600; // Default Baud Rate for entities
}
```

**Figure 14: Global tags in GridSim package.**

`class gridsim.ResGridlet` – It represents a Gridlet submitted to the resource for processing. It contains Gridlet object along with its arrival time and the ID of machine and PE allocated to it. It acts as a placeholder for maintaining the amount of resource share allocated at various times for simulating time-shared scheduling using internal events.

`class gridsim.GridStatistics` – This is a GridSim entity that records statistical data reported by other entities. It stores data objects with their label and timestamp. At the end of simulation, the user-defined report-writer entity can query recorded statistics of interest for report generation.

`class gridsim.Accumulator` – The objects of this class provide a placeholder for maintaining statistical values of a series of data added to it. It can be queried for mean, sum, standard deviation, and the largest and smallest values in the data series.

`class gridsim.GridSimShutdown` – This is a GridSim entity that waits for termination of all User entities to determine the end of simulation. It then signals the user-defined report-writer entity to interact with GridStatistics entity to generate report. Finally, it signals the end of simulation to other GridSim core entities.

`class gridsim.GridSimRandom` – This class provides static methods for incorporating randomness in data used for any simulation. Any predicted/estimated data, e.g., number of Gridlets used by an experiment, execution time and output size of a Gridlet etc., need to be mapped to real-world data by introducing randomness to reflect the uncertainty that is present in the prediction/estimation process and the randomness that exists in the nature itself. The execution time of a Gridlet on a particular resource, for example, can vary depending on the local load, which is not covered by the scope of GridSim to simulate.

The real($d$, $f_L$, $f_M$) method of this class maps the predicted/estimated value $d$ to a random real-world value between $(1- f_L)\times d$ to $(1+ f_M)\times d$, using the formula $d\times(1 - f_L + (f_L + f_M)\times rd)$ where $0.0 \le f_L, f_M \le 1.0$ and $rd$ is a uniformly distributed `double` value between 0.0 and 1.0. This class also maintains different values of $f_L$ and $f_M$ factors for different situations to represent different level of uncertainty involved.

# 4 Building Simulations with GridSim

To simulate grid resource brokers using the GridSim toolkit, the developers need to create new entities that exhibit the behavior of grid users and scheduling systems. The user-defined entities extend the GridSim base class to inherit the properties of concurrent entities capable of communicating with other entities using events. The detailed steps involved in modeling resources and applications, and simulating brokers using the GridSim toolkit are discussed below. We then present the simulation of a Nimrod-G like resource broker that implements deadline and budget constrained scheduling algorithms.

## 4.1 A Recipe for Simulating Application Scheduling

In this section we present high-level steps, with sample code clips, to demonstrate how GridSim can be used to simulate a Grid environment to analyze scheduling algorithms:

- First, we need to create Grid resources of different capability and configuration (a single or multiprocessor with time/space-shared resource manager) similar to those present in the World-Wide Grid (WWG) testbed [11]). We also need to create users with different requirements (application and quality of service requirements). A sample code for creating grid environment is given in Figure 15.

```
public static void CreateSampleGridEnvironement(int no_of_users, int no_of_resources,
  double B_factor, double D_factor, int policy, double how_long, double seed) {
  Calendar now = Calendar.getInstance();

  String ReportWriterName = "MyReportWriter";
  GridSim.Init(no_of_users, calender, true, eff, efp, ReportWriterName);

  String[] category = {"*.USER.TimeUtilization", "*.USER.GridletCompletionFactor",
"*.USER.BudgetUtilization"};

// Create Report Writer Entity and category indicates types of information to be recorded.
  new ReportWriter(ReportWriterName, no_of_users, no_of_resources, ReportFile, category,
report_on_next_row_flag);

  // Create Resources
  for(int i=0; i<no_of_resources; i++) {
     // Create PEs
     PEList peList = new PEList();
     for(int j=0; j<(i*1+1); j++)
        peList.add(new PE(0, 100));

     // Create machine list
     MachineList mList = new MachineList();
     mList.add(new Machine(0, peList));

     // Create a resource containing machines
     ResourceCharacteristics resource = new ResourceCharacteristics("INTEL", "Linux",
         mList, ResourceCharacteristics.TIME_SHARED, 0.0, i*0.5+1.0);
     LinkedList Weekends = new LinkedList();
     Weekends.add(new Integer(Calendar.SATURDAY));
     Weekends.add(new Integer(Calendar.SUNDAY));
     LinkedList Holidays = new LinkedList(); // no holiday is set!

     // Setup resource as simulated entity with a name (e.g. "Resource_1").
     new GridResource("Resource_"+i, 28000.0, seed, resource,
                                          0.0, 0.0, 0.0, Weekends, Holidays);
  }
  Random r = new Random(seed);
  // Create Application, Experiment, and Users
  for(int i=0; i<no_of_users; i++)
  {
    Random r = new Random(seed*997*(1+i)+1);
    GridletList glList = Application1(r);  // it creates Gridlets and returns their list
    Experiment expt = new Experiment(0, glList, policy, true, B_factor, D_factor);
    new UserEntity("U"+i, expt, 28000.0, how_long, seed*997*(1+i)+1, i, user_entity_report);
  }
  // Perform Simulation
  GridSim.Start();
}
```

**Figure 15: A sample code segment for creating Grid resource and user entities in GridSim.**

- Second, we need to model applications by creating a number of Gridlets (that appear similar to Nimrod-G jobs) and define all parameters associated with jobs as shown in Figure 16. The Gridlets need to be grouped together depending on the application model.

```
Gridlet gl = new Gridlet(Gridlet_id, Gridlet_length, GridletFileSize,
                       GridletOutputSize);
```

**Figure 16: The Gridlet method in GridSim.**

- Then, we need to create a GridSim User entity that creates and interacts with the resource broker scheduling entity to coordinate execution experiment. It can also directly interact with GIS and resource entities for Grid information and submitting or receiving processed Gridlets, however, for modularity sake, we encourage the implementation of a separate resource broker entity by extending the GridSim class.
- Finally, we need to implement a resource broker entity that performs application scheduling on Grid resources. A sample code for implementing the broker is shown in Figure 17. First, it inquires the Grid Information Service (GIS), and then inquires for resource capability including cost. Depending on processing requirements, it develops schedule for assigning Gridlets to resources and coordinates the execution. The scheduling policies can be systems-centric like those implemented in many Grid systems such as Condor or user-centric like the Nimrod-G broker's quality of service (QoS) driven application scheduling algorithms [10].

```
class Broker extends GridSim {
  private Experiment experiment;
  private LinkedList ResIDList;
  private LinkedList BrokerResourceList;

  public Broker(String name, double baud_rate)
  {
    super(name, baud_rate);
    GridletDispatched = 0;
    GridletReturned = 0;
    Expenses = 0.0;
    MaxGridletPerPE = 2;
  }

  ... // Gridlet scheduling flow code at the Grid Resource Broker level

  public void body() {

    Sim_event ev = new Sim_event();
    // Accept User Commands and Process
    for( sim_get_next(ev); ev.get_tag()!=GridSimTags.END_OF_SIMULATION; sim_get_next(ev))
    {
      experiment = (Experiment) ev.get_data();
      int UserEntityID = ev.get_src();

      // Record Experiment Start Time.
      experiment.SetStartTime();

     // Set Gridlets' OwnerID as this BrokerID so that Resources knows where to return them.
      for(int i=0; i<experiment.GetGridletList().size(); i++)
        ((Gridlet) experiment.GetGridletList().get(i)).SetUserID(get_id());

      // RESOURCE DISCOVERY
      ResIDList = (LinkedList) GetGridResourceList();

      // RESOURCE TRADING and SORTING
      // SCHEDULING
      while (glFinishedList.size() < experiment.GetGridletList().size())
      {
        if((GridSim.Clock()>=experiment.GetDeadline())||(Expenses>=experiment.GetBudget()) )
          break;

        scheduled_count = ScheduleAdviser();
        dispatched_count = Dispatcher();
        received_count = Receiver();

        // Heurisitics for deciding hold condition
        if(dispatched<=0 && received<=0 && glUnfinishedList.size()>0)
        {
          double deadline_left = experiment.GetDeadline()-GridSim.Clock();
          GridSimHold(Math.max(deadline_left*0.01, 1.0));
        }
      }
    }
  ...  // Code for actual scheduling policy
  ...  // Code for dispatch policy
  }
 }
```

**Figure 17: A sample code segment for creating a Grid resource broker in GridSim.**

## 4.2 Economic Grid Resource Broker Simulation

We used the GridSim toolkit to simulate grid environment and a Nimrod-G like deadline and budget constrained scheduling system called economic grid resource broker. The simulated grid environment contains multiple resources and user entities with different requirements. The users create an experiment that contains application specification (a set of Gridlets that represent application jobs with different processing) and quality of service requirements (deadline and budget constraints with optimization strategy). We created two entities that simulate users and the brokers by extending the GridSim class. When simulated, each user entity having its own application and quality of service requirements creates its own instance of the broker entity for scheduling Gridlets on resources.

### 4.2.1 Broker Architecture

The broker entity architecture along with its interaction flow diagram with other entities is shown in Figure 18. The key components of the broker are: experiment interface, resource discovery and trading, scheduling flow manager backed with scheduling heuristics and algorithms, Gridlets dispatcher, and Gridlets receptor.

The following high-level steps describe functionality of the broker components and their interaction:

1. The user entity creates an experiment that contains application description (a list of Gridlets to be processed) and user requirements to the broker via the experiment interface.
2. The broker resource discovery and trading module interacts with the GridSim GIS entity to identify contact information of resources and then interacts with resources to establish their configuration and access cost. It creates a Broker Resource list that acts as placeholder for maintaining resource properties, a list of Gridlets committed for execution on the resource, and the resource performance data as predicted through the measure and extrapolation methodology.
3. The scheduling flow manager selects an appropriate scheduling algorithm for mapping Gridlets to resources depending on the user requirements. Gridlets that are mapped to a specific resource are added to the Gridlets list in the Broker Resource.
4. For each of the resources, the dispatcher selects the number of Gridlets that can be staged for execution according to the usage policy to avoid overloading resources with single user jobs.
5. The dispatcher then submits Gridlets to resources using the GridSim's asynchronous service.
6. When the Gridlet processing completes, the resource returns it to the broker's Grid let receptor module, which then measures and updates the runtime parameter, *resource or MI share available to the user*. It aids in predicting the job consumption rate for making scheduling decisions.
7. The steps, 3–6, continue until all the Gridlets are processed or the broker exceeds deadline or budget limits. The broker then returns updated experiment data along with processed Gridlets back to the user entity.

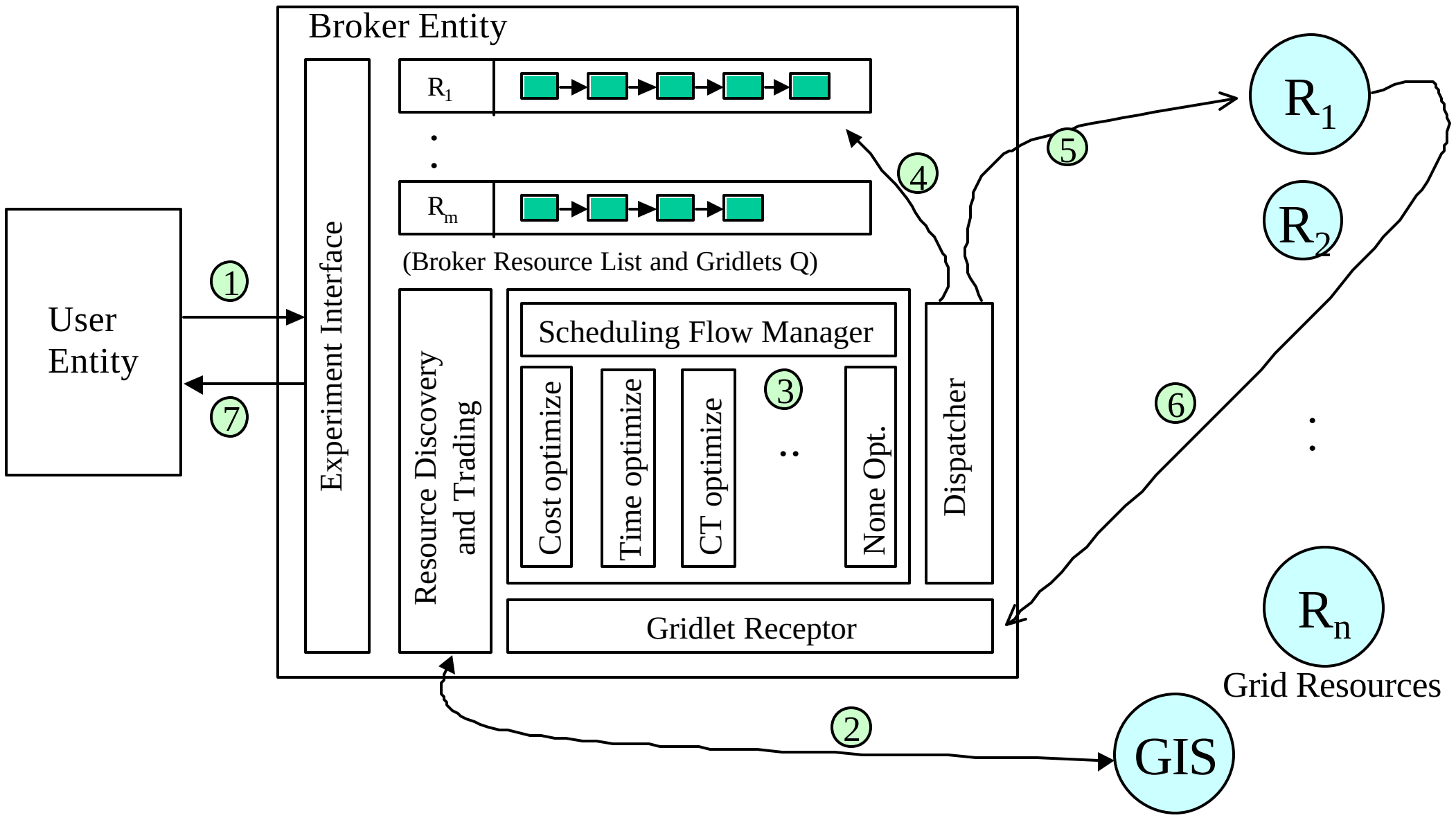

**Figure 18: Economic grid resource broker architecture and its interaction with other entities.**

A class diagram hierarchy of the grid broker package built using the GridSim toolkit is shown in Figure 19. The grid broker package implements the following key classes:

`class Experiment` – It acts as a placeholder for representing simulation experiment configuration that includes synthesized application (a set of Gridlets stored in GridletList) and user requirements such as D and B-factors or deadline and budget constraints, and optimization strategy. It provides methods for updating and querying the experiment parameters and status. The user entity invokes the broker entity and passes its requirements via experiment object. On receiving an experiment from its user, the broker schedules Gridlets according to the optimization policy set for the experiment.

`class UserEntity` – A GridSim entity that simulates the user. It invokes the broker and passes the user requirements. When it receives the results of application processing, it records parameters of interest with the gridsim.Statistics entity. When it has no more processing requirements, it sends `END_OF_SIMULATION` event to the gridsim.GridSimShutdown entity.

`class Broker` – A GridSim entity that simulates the Grid resource broker. On receiving an experiment from the user entity, it does resource discovery, and determines deadline and budget values based on D and B factors, and then proceeds with scheduling. It schedules Gridlets on resources depending on user constraints, optimization strategy, and cost of resources and their availability. When it receives the results of application processing, it records parameters of interest with the gridsim.Statistics entity. When it has no more processing requirements, it sends `END_OF_SIMULATION` event to the gridsim.GridSimShutdown entity. The interaction between the broker and other GridSim entities is shown in Figure 5 for time-shared resources and Figure 6 for space-shared resources.

`class BrokerResource` – It acts as placeholder for the broker to maintain a detailed record of the resources it uses for processing user application. It maintains resource characteristics, a list of Gridlets assigned to the resource, the actual amount of MIPS available to the user, and a report on the Gridlets processed. These measurements help in extrapolating and predicting the resource performance from the user point of view and aid in scheduling jobs dynamically at runtime.

`class ReportWriter` – A user-defined, optional GridSim entity which is meant for creating a report at the end of each simulation by interacting with the gridsim.Statistics entity. If the user does not want to create a report, then can pass "`null`" as the name of ReportWriter entity. Note that users can choose any name for the ReportWriter entity and for the class name since all entities are identified by their name defined at the runtime.

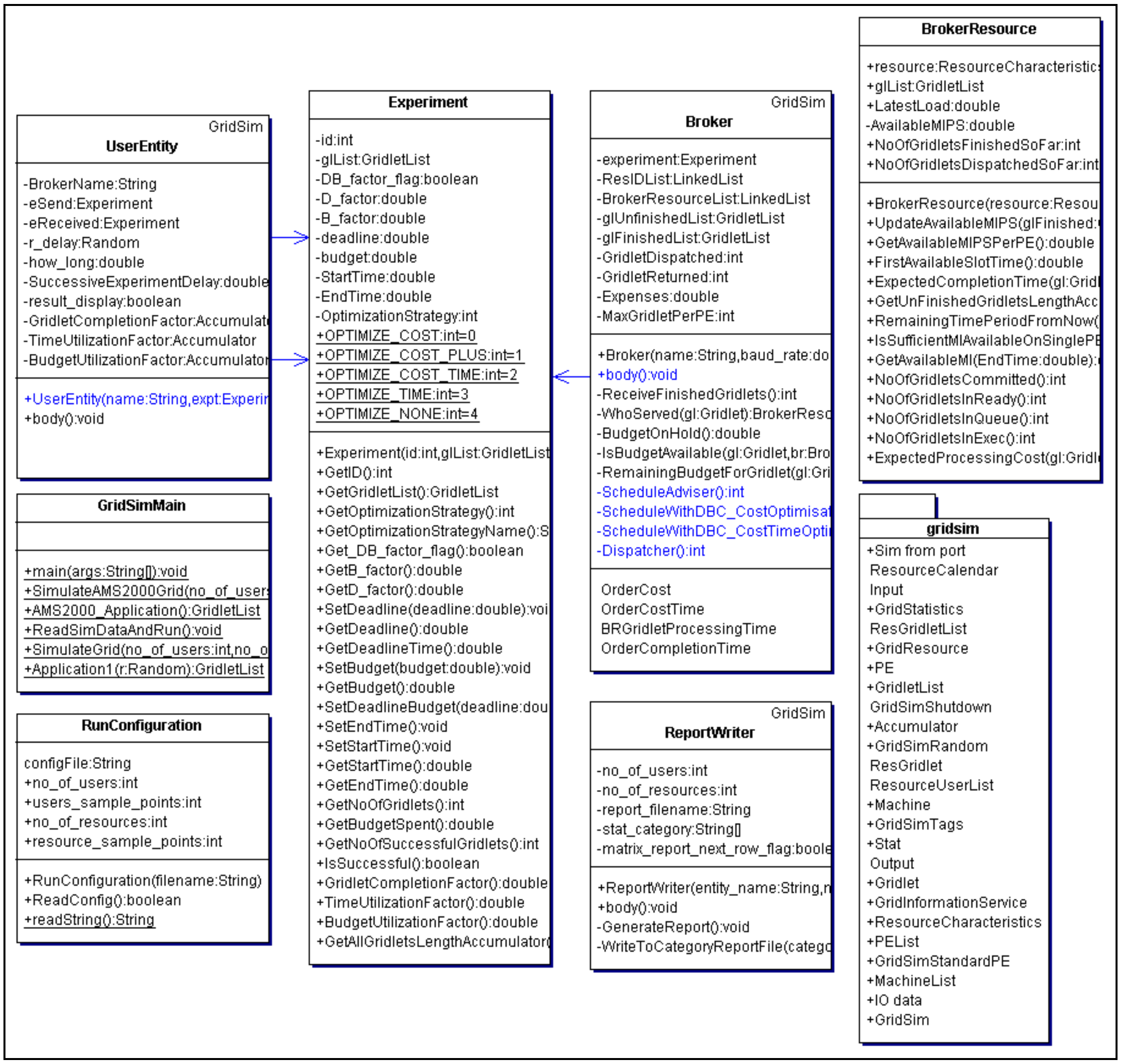

**Figure 19: A class hierarchy diagram of Grid broker using the gridsim package.**

### 4.2.2 Deadline and Budget Constrained Cost Optimisation Scheduling Algroithm

We have simulated deadline and budget constrained (DBC) scheduling algorithms, cost-optimisation, time-optimisation, and none-optimisation, presented in [10]. We have also proposed a new scheduling algorithm, called cost-time optimisation [23], which is a variant of cost and time optimisation algorithms and simulated its performance.

The steps for implementing DBC cost-optimisation scheduling algorithms within economic broker simulator are shown in Figure 20. This algorithm attempts to process jobs as economically as possible within the deadline and budget. The results of simulation are discussed in the next section.

*Algorithm: DBC_Scheduling_with_Cost_Optimisation()*

1. RESOURCE DISCOVERY: Identify resources that can be used in this execution with their capability through the Grid Information Service.
2. RESOURCE TRADING: Identify cost of each of the resources in terms of CPU cost per second and capability to be delivered per cost-unit.
3. If the user supplies D and B factors, then determine the absolute deadline and budget based on the capability and cost of resources and user's requirements.
4. SORT resources by increasing order of cost.
5. SCHEDULING: Repeat while there exists unprocessed jobs in application job list with a delay of scheduling event period or occurrence of an event AND the time and process expenses are within deadline and budget limits:

   [SCHEDULE ADVISOR with Policy]

   a. For each resource predict and establish the job consumption rate or the available resource share through measure and extrapolation.
   b. For each resource based on its job consumption rate or available resource share, predict and establish the number of jobs a resource can process by the deadline.
   c. For each resource in order:
      i. If the number of jobs currently assigned to a resource is less than the predicted number of jobs that a resource can consume, assign more jobs from unassigned job queue or from the most expensive machines based on job state and feasibility. Assign job to a resource only when there is enough budget available.
      ii. Alternatively, if a resource has more jobs than it can complete by the deadline, move those extra jobs to unassigned job queue.

   [DISPATCHER with Policy]

   d. The dispatcher takes care of submission of jobs to remote machines with submission and resource policy and constraints depending on resource type (time or space shared).

**Figure 20: Deadline and budget constrained (DBC) scheduling with cost-optimization.**

### 4.2.3 Determining the Deadline and Budget

A D-factor close to 1 signifies the user's willingness to set highly relaxed deadline, which is sufficient to process application even when only the slowest resources are available. Similarly a B-factor close to 1 signifies that the user is willing to spend as much money as required even when only the most expensive resource is used. The user jobs are scheduled on the Grid through its own broker. The broker uses these factors in determining the absolute deadline (see Equation 1) and budget (see Equation 2) values for a given execution scenario at runtime as follows:

***Computing the Deadline:***

$$Deadline = T_{MIN} + D_{FACTOR} * (T_{MAX} - T_{MIN})$$

**Equation 1**

where,

- $T_{MIN}$ = the time required to process all the jobs, in parallel, giving the fastest resource the highest priority.
- $T_{MAX}$ = the time required to process all the jobs, serially, using the slowest resource.
- An application with $D_{FACTOR}$ **< 0** would **never** be completed.
- An application with $D_{FACTOR} \geq$ **1** would **always** be completed as long as some resources are available throughout the deadline.

***Computing the Budget:***

$$Budget = C_{MIN} + B_{FACTOR} * (C_{MAX} - C_{MIN})$$

**Equation 2**

where,

- $C_{MIN}$ = the cost of processing all the jobs, in parallel within deadline, giving the cheapest resource the highest priority.
- $C_{MAX}$ = the cost of processing all the jobs, in parallel within deadline, giving the costliest resource the highest priority.
- An application with $B_{FACTOR}$ **< 0** would **never** be completed.
- An application with $B_{FACTOR} \geq$ **1** would **always** be completed as long as some resources are available throughout the deadline.

# 5 Scheduling Simulation Experiments

To simulate application scheduling in GridSim environment using the economic grid broker requires the modeling and creation of GridSim resources and applications that model jobs as Gridlets. In this section, we present resource and application modeling along with the results of experiments with quality of services driven application processing.

## 5.1 Resource Modeling

We modeled and simulated a number of time- and space-shared resources with different characteristics, configuration, and capability as those in the WWG testbed. We have selected the latest CPUs models AlphaServer ES40, Sun Netra 20, Intel VC820 (800EB MHz, Pentium III), and SGI Origin 3200 1X 500MHz R14k released by their manufacturers Compaq, Sun, Intel, and SGI respectively. The processing capability of these PEs in simulation time-unit is modeled after the base value of SPEC CPU (INT) 2000 benchmark ratings published in [22]. To enable the users to model their application processing requirements, we *assumed the MIPS rating of PEs same as the SPEC rating*.

Table 2 shows characteristics of resources simulated and their PE cost per time unit in G$ (Grid dollar). These simulated resources resemble the WWG testbed resources used in processing a parameter sweep application using the Nimrod-G broker [24]. The PE cost in G$/unit time not necessarily reflects the cost of processing when PEs have different capability. The brokers need to translate it into the G$ per MI (million instructions) for each resource. Such translation helps in identifying the relative cost of resources for processing Gridlets on them.

**Table 2: WWG testbed resources simulated using GridSim.**

| Resource Name in Simulation | Simulated Resource Characteristics Vendor, Resource Type, Node OS, No of PEs | Equivalent Resource in Worldwide Grid (Hostname, Location) | A PE SPEC/ MIPS Rating | Resource Manager Type | Price (G$/PE time unit) | MIPS per G$ |
|---|---|---|---|---|---|---|
| R0 | Compaq, AlphaServer, CPU, OSF1, 4 | grendel.vpac.org, VPAC, Melb, Australia | 515 | Time-shared | 8 | 64.37 |
| R1 | Sun, Ultra, Solaris, 4 | hpc420.hpcc.jp, AIST, Tokyo, Japan | 377 | Time-shared | 4 | 94.25 |
| R2 | Sun, Ultra, Solaris, 4 | hpc420-1.hpcc.jp, AIST, Tokyo, Japan | 377 | Time-shared | 3 | 125.66 |
| R3 | Sun, Ultra, Solaris, 2 | hpc420-2.hpcc.jp, AIST, Tokyo, Japan | 377 | Time-shared | 3 | 125.66 |
| R4 | Intel, Pentium/VC820, Linux, 2 | barbera.cnuce.cnr.it, CNR, Pisa, Italy | 380 | Time-shared | 2 | 190.0 |
| R5 | SGI, Origin 3200, IRIX, 6 | onyx1.zib.de, ZIB, Berlin, Germany | 410 | Time-shared | 5 | 82.0 |
| R6 | SGI, Origin 3200, IRIX, 16 | Onyx3.zib.de, ZIB, Berlin, Germany | 410 | Time-shared | 5 | 82.0 |
| R7 | SGI, Origin 3200, IRIX, 16 | mat.ruk.cuni.cz, Charles U., Prague, Czech Republic | 410 | Space-shared | 4 | 102.5 |
| R8 | Intel, Pentium/VC820, Linux, 2 | marge.csm.port.ac.uk, Portsmouth, UK | 380 | Time-shared | 1 | 380.0 |
| R9 | SGI, Origin 3200, IRIX, 4 (accessible) | green.cfs.ac.uk, Manchester, UK | 410 | Time-shared | 6 | 68.33 |
| R10 | Sun, Ultra, Solaris, 8, | pitcairn.mcs.anl.gov, ANL, Chicago, USA | 377 | Time-shared | 3 | 125.66 |

## 5.2 Application Modeling

We have modeled a task farming application that consists of 200 jobs. In GridSim, these jobs are packaged as Gridlets whose contents include the job length in MI, the size of job input and output data in bytes along with various other execution related parameters when they move between the broker and resources. The job length is expressed in terms of the time it takes to run on a standard resource PE with SPEC/MIPS rating of 100. Gridlets processing time is expressed in such a way that they are expected to take at least 100 time-units with a random variation of 0 to 10% on the positive side of the standard resource. That means, Gridlets' job length (processing requirements) can be at least 10,000 MI with a random variation of 0 to 10% on the positive side. This 0 to 10% random variation in Gridlets' job length is introduced to model heterogeneous tasks similar to those present in the real world parameter sweep applications.

## 5.3 DBC Scheduling Experiments with Cost-Optimization—for a Single User

In this experiment, we performed scheduling experiments with different values of deadline and budget constraints (DBC) for a single user. The deadline is varied in simulation time from 100 to 3600 in steps of 500. The budget is varied from G$ 5000 to 22000 in steps of 1000. For this scenario, we performed scheduling simulation for DBC *cost-optimization* algorithm. The number of Gridlets processed, deadline utilized, and budget spent for different scheduling scenario is shown in Figure 21–Figure 24. From

Figure 21, it can be observed that for a tight deadline (e.g., 100 time unit), the number of Gridlets processed increases as the budget value increases. Because, when a higher budget is available, the broker leases expensive resources to process more jobs within the deadline. Alternatively, when scheduling with a low budget value, the number of Gridlets processed increases as the deadline is relaxed (see Figure 22).

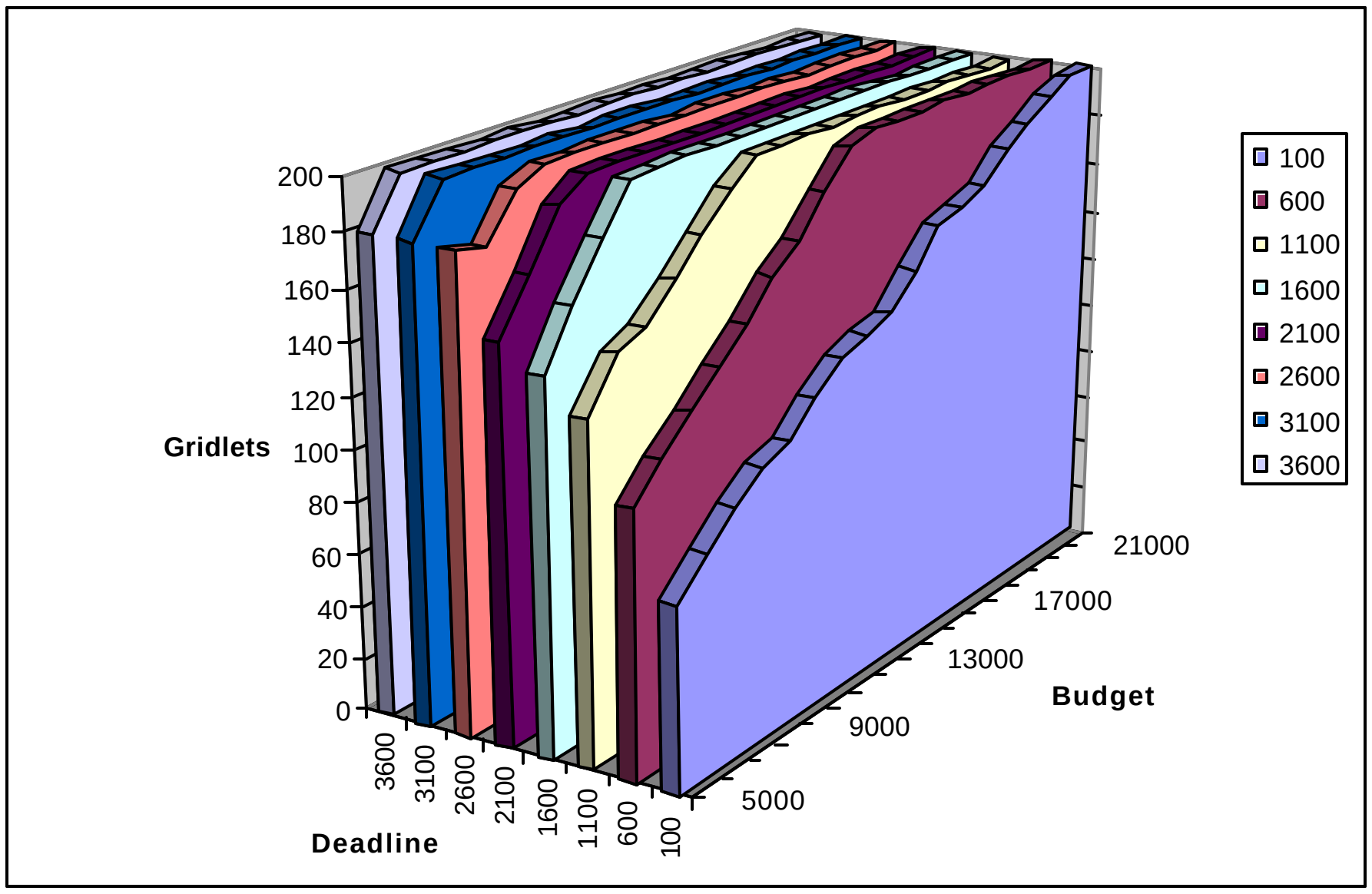

**Figure 21: No. of Gridlets processed for different budget limits with a fixed deadline for each.**

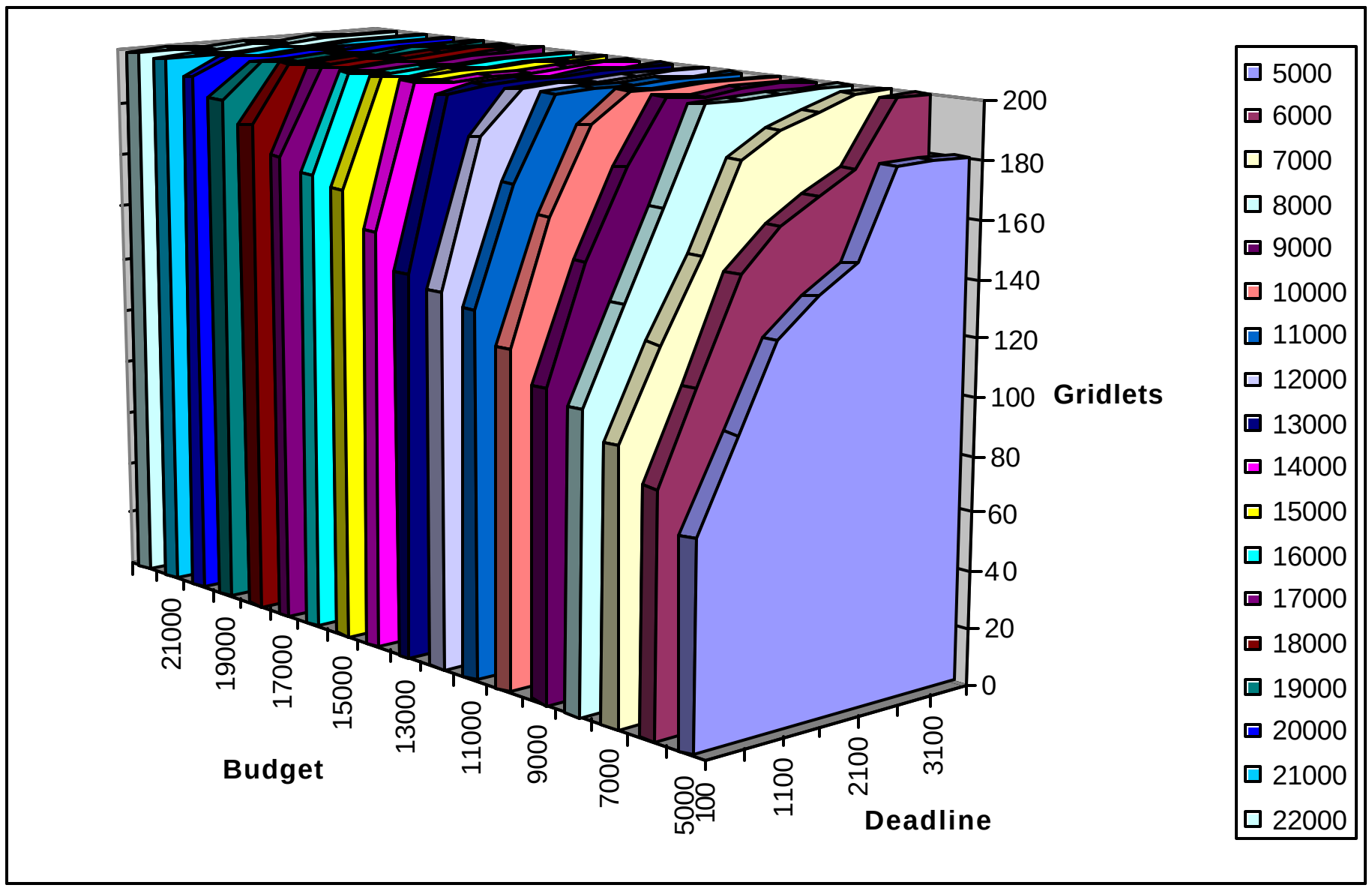

**Figure 22: No. of Gridlets processed for different deadline limits with a fixed budget for each.**

The impact of budget for different values of deadline is shown in Figure 23. In cost-optimization scheduling, for a larger deadline value (see time utilization for deadline of 3600), the increase in budget value does not have much impact on resource selection. This trend can also be observed from the budget spent for processing Gridlets with different deadline constraints (see Figure 24). When the deadline is too tight (e.g., 100), it is likely that the complete budget is spent for processing Gridlets within the deadline.

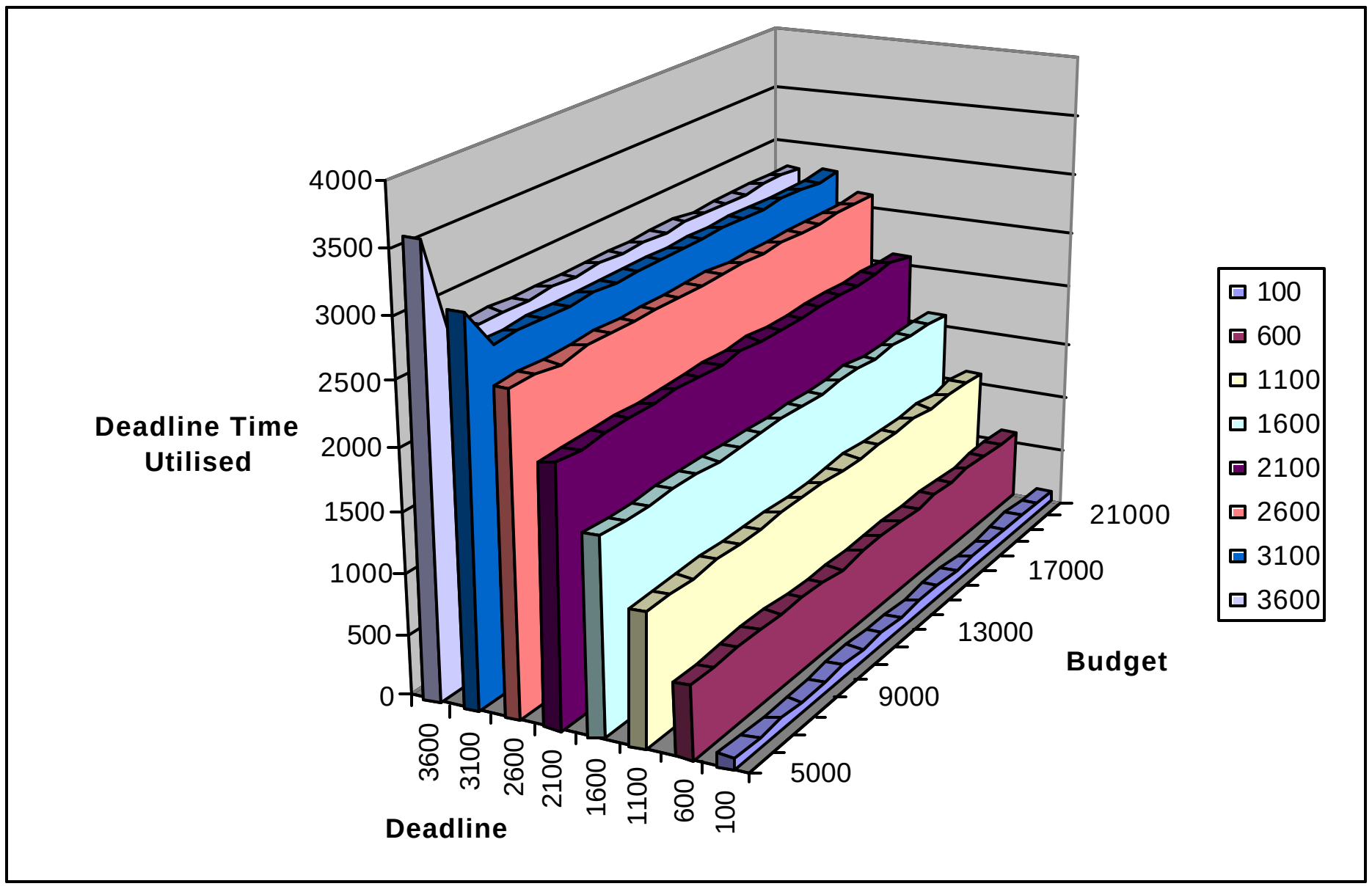

**Figure 23: Deadline time utilized for processing Gridlets for different values of deadline and budget.**

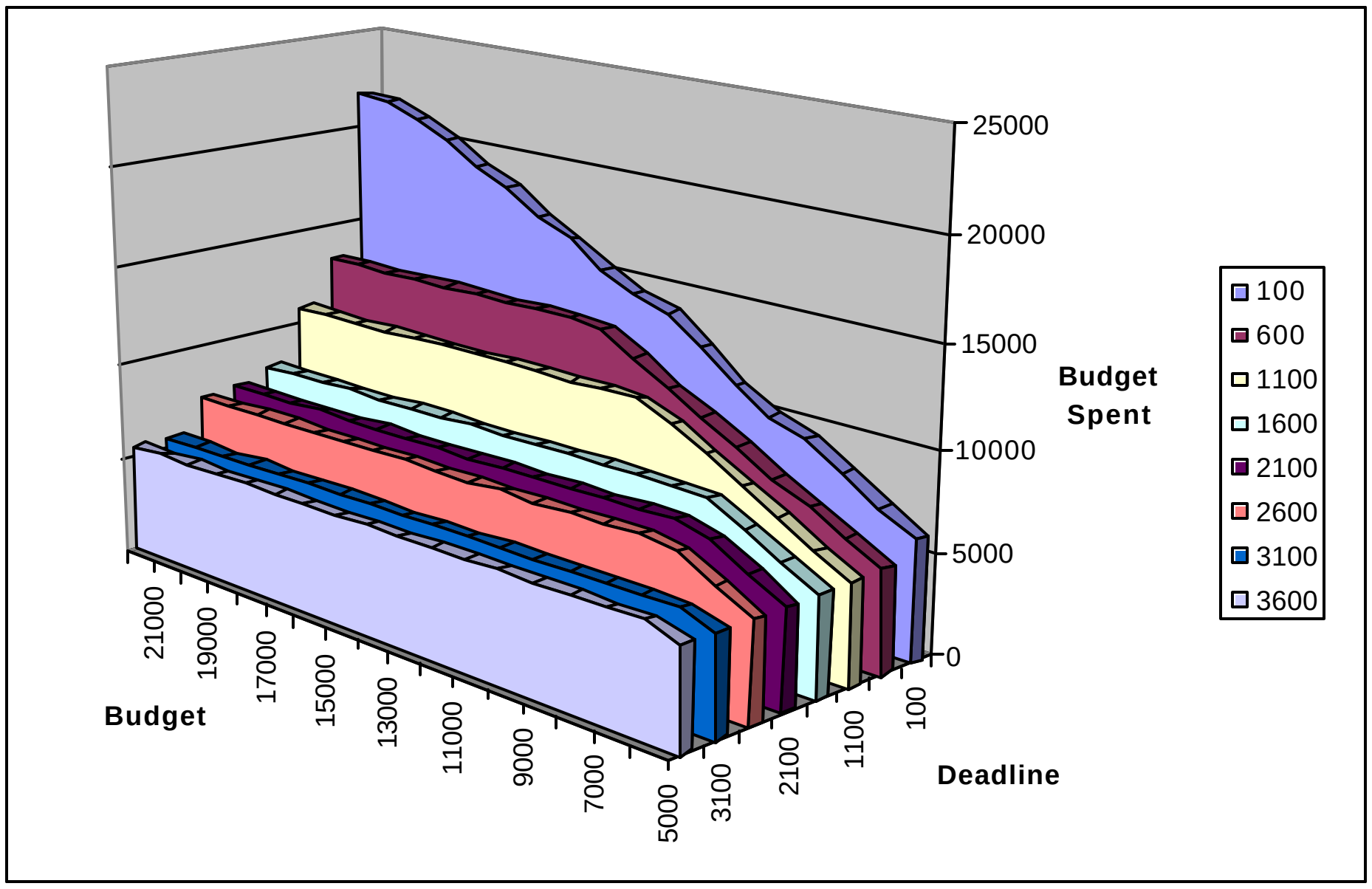

**Figure 24: Budget spent for processing Gridlets for different values of deadline and budget.**

Three diagrams (Figure 25–Figure 27) show the selection of resources for processing Gridlets for different budget values with a fixed deadline of 100, 1100, and 3100 (low, medium, and high deadline value) respectively. It can be observed that when the deadline is low, the economic broker also leases expensive resources to process Gridlets whenever the budget permits (see Figure 25). In this, all resources have been used depending on the budget availability. When the deadline is increased to a high value (a medium deadline of 1100), the broker processes as many Gridlets as possible on cheaper resources by the deadline (see Figure 26) and utilizes expensive resources if required. When the deadline is highly relaxed (a high deadline of 3100), the broker allocated Gridlets to the cheapest resource since it was able to process all Gridlets within this deadline (see Figure 27). In all three diagrams (Figure 25 –Figure 27), the left most solid curve marked with the label "All" in the resources axis represents the aggregation of all resources and shows the total number of Gridlets processed for the different budgets.

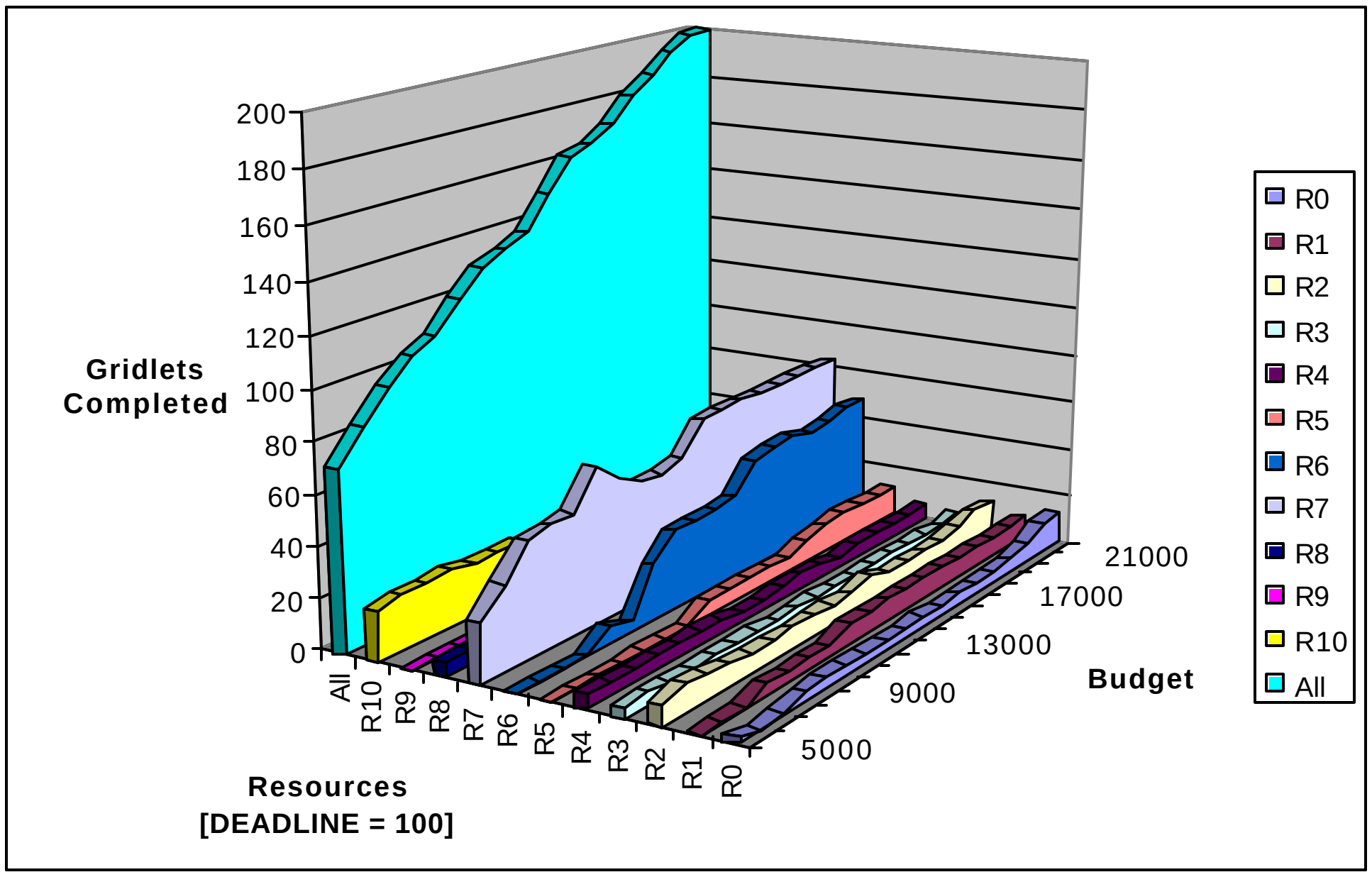

**Figure 25: Gridlets processed on resources for different budget values with low deadline.**

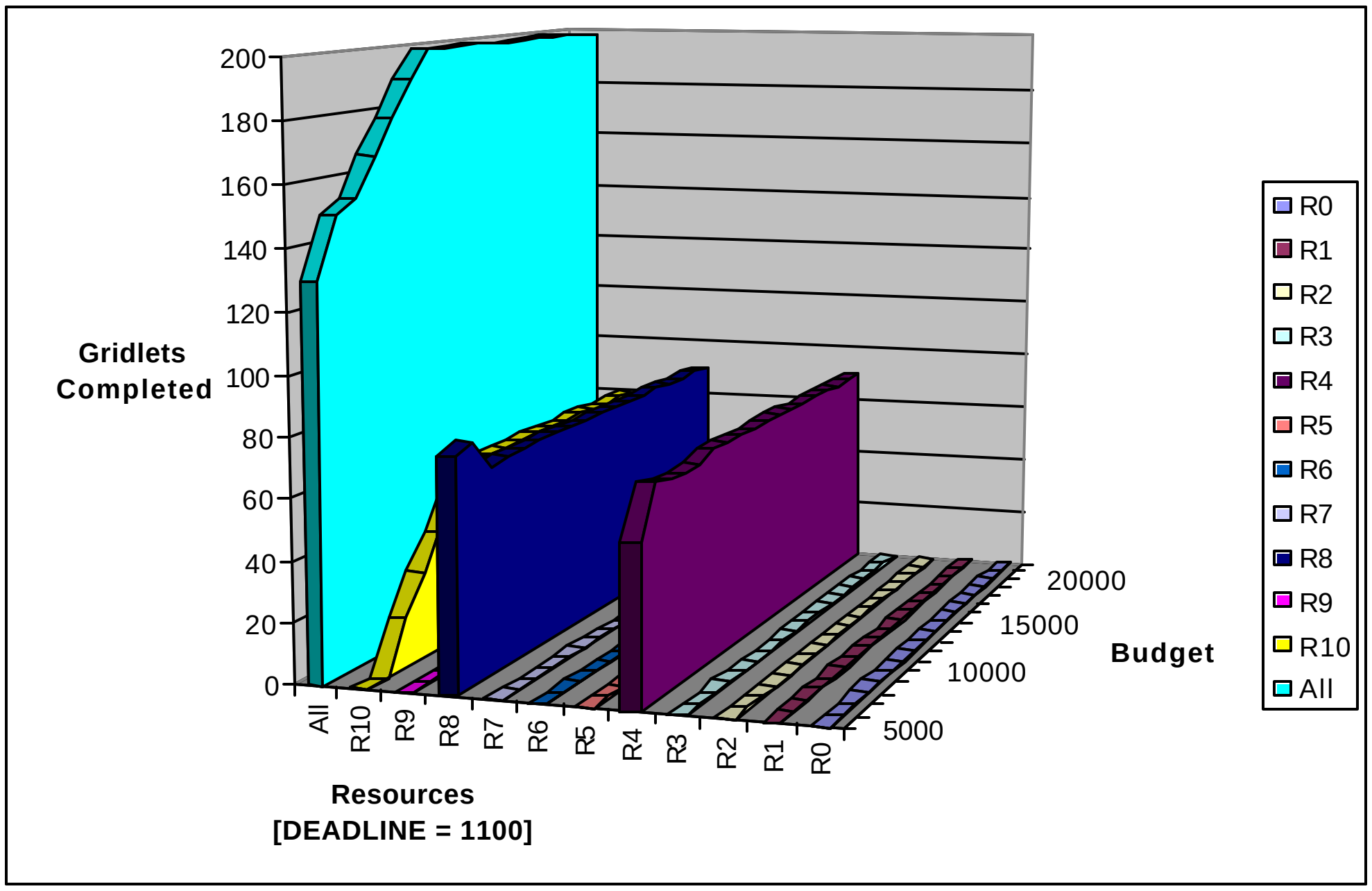

**Figure 26: Gridlets processed on resources for different budget values with medium deadline.**

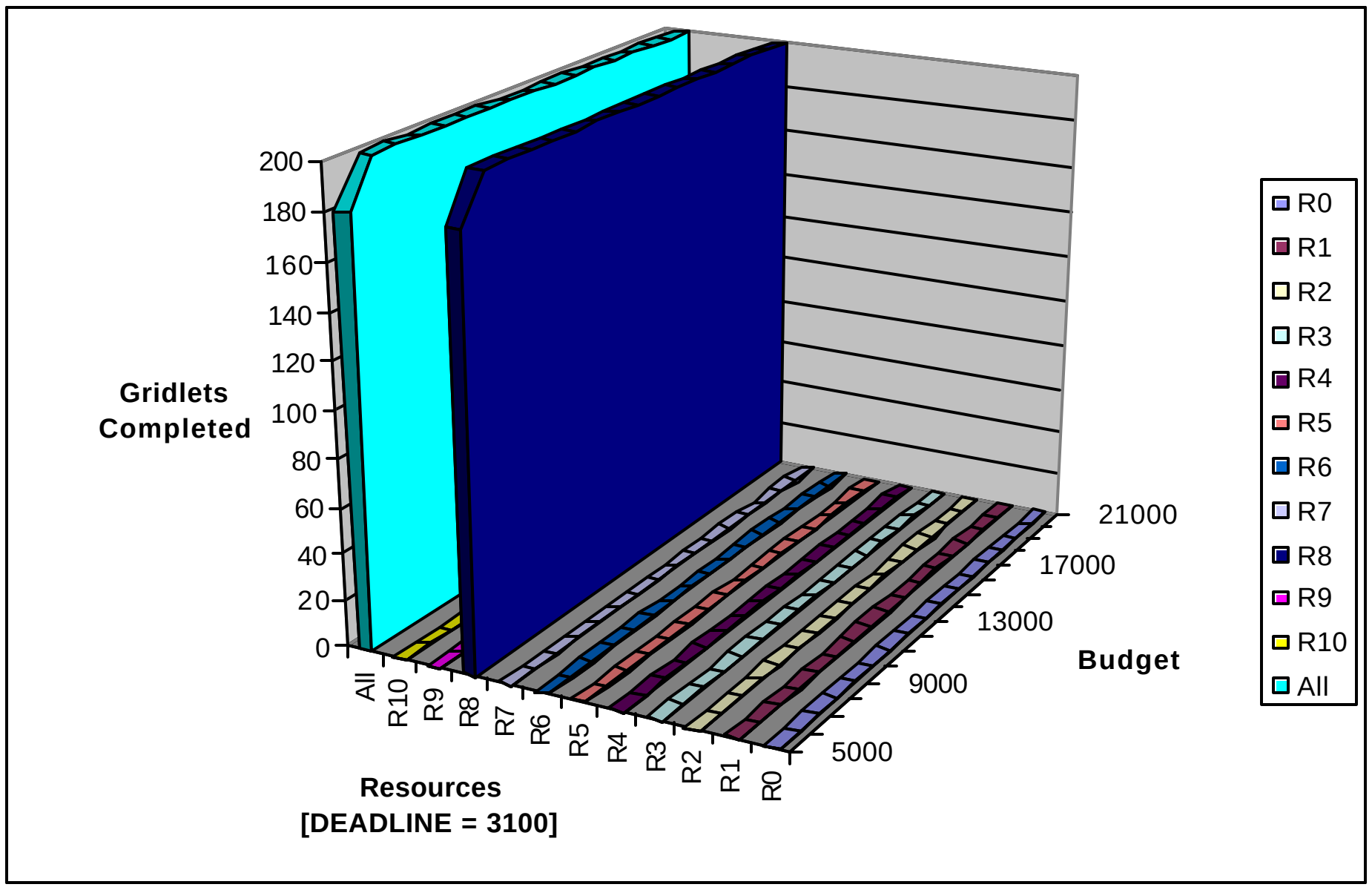

**Figure 27: Gridlets processed on resources for different budget values with high deadline.**

Let us now take a microscopic look at the allocation of resources at different times during the scheduling experimentation. The two graphs (Figure 28 and Figure 30) show a trace of leasing resources at different times during the scheduling experiment for processing Gridlets for different budget values with a fixed deadline of 100 and 3100 (low and high deadline value) respectively. It can be observed that when the deadline value is low, the economic broker also leases expensive resources to process Gridlets whenever the budget permits. The broker had to allocate powerful resources even if they are expensive since the deadline is too tight (see Figure 28 for Gridlets completed and Figure 29 for budget spent in processing). But this is not the case when the deadline is highly relaxed (see Figure 30)—the broker leased just one resource, which happened to process all Gridlets within the given deadline. From the diagrams (Figure 28 and Figure 29), it can be observed that the resource R7 has processed more Gridlets than the resource R6, but had to spent more budget on the resource R6 since it is more expensive than the resource R7.

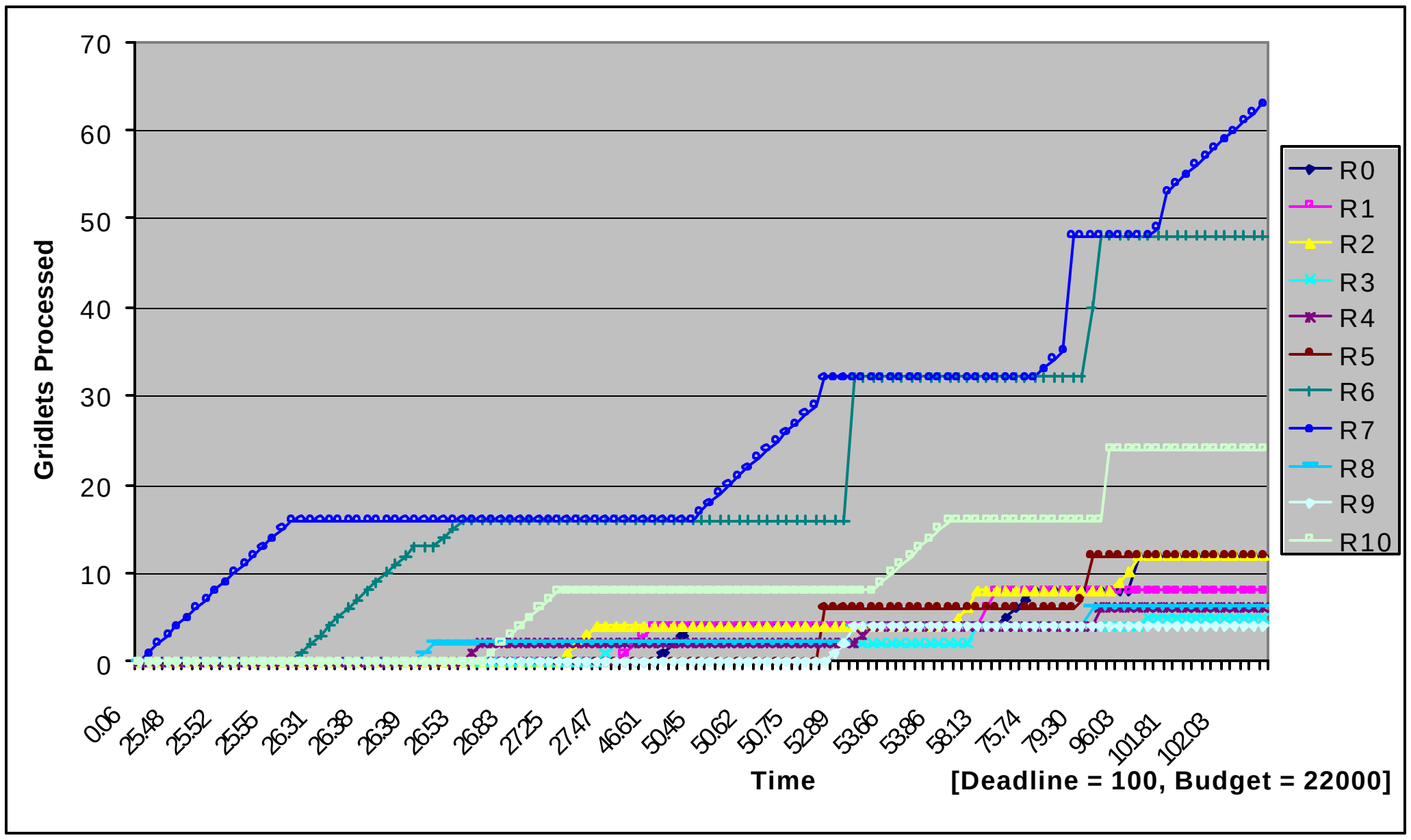

**Figure 28: Trace of No. of Gridlets processed on resources for a low deadline and high budget constraints.**

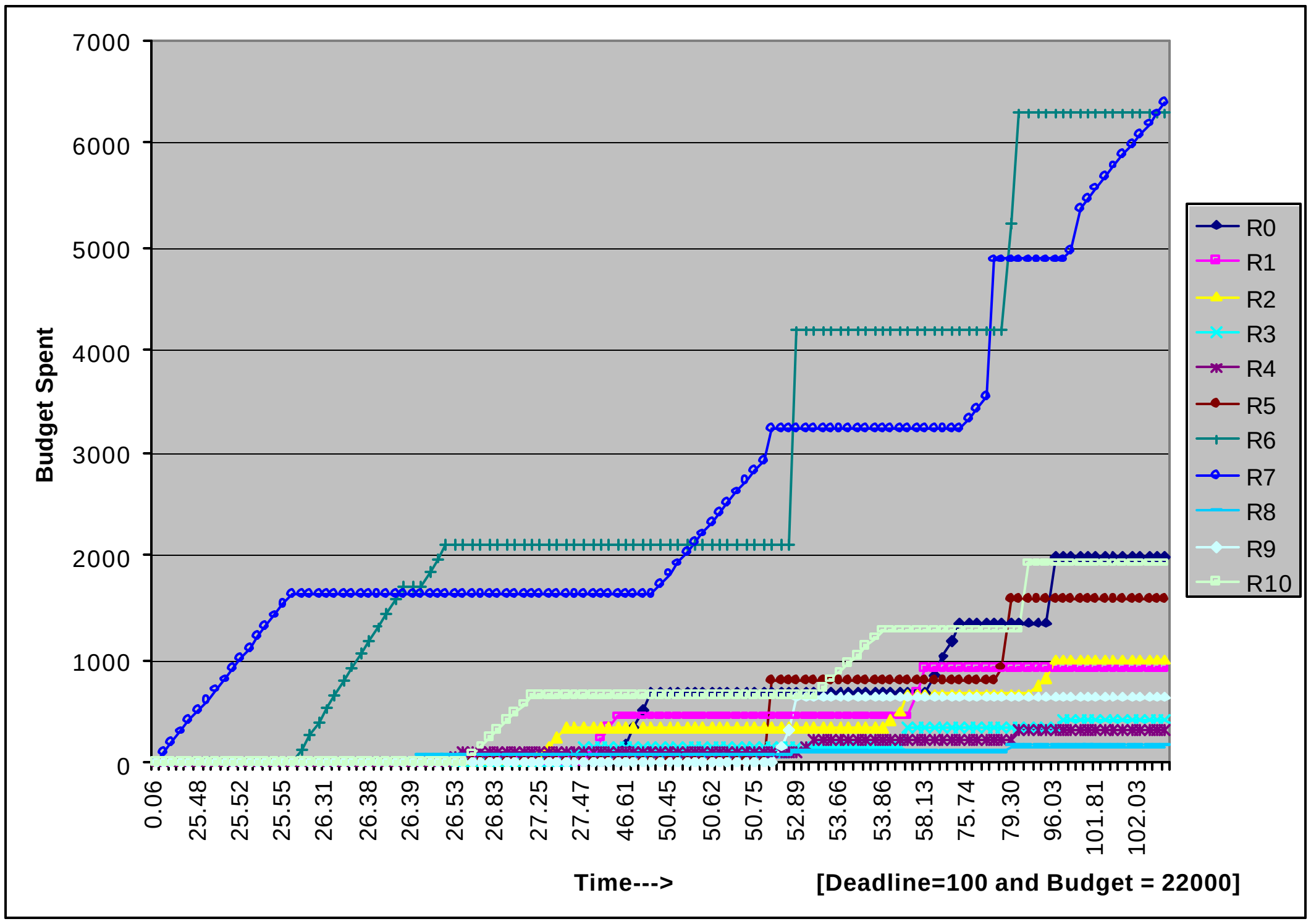

**Figure 29: Trace of budget spent for low deadline and high budget constraints.**

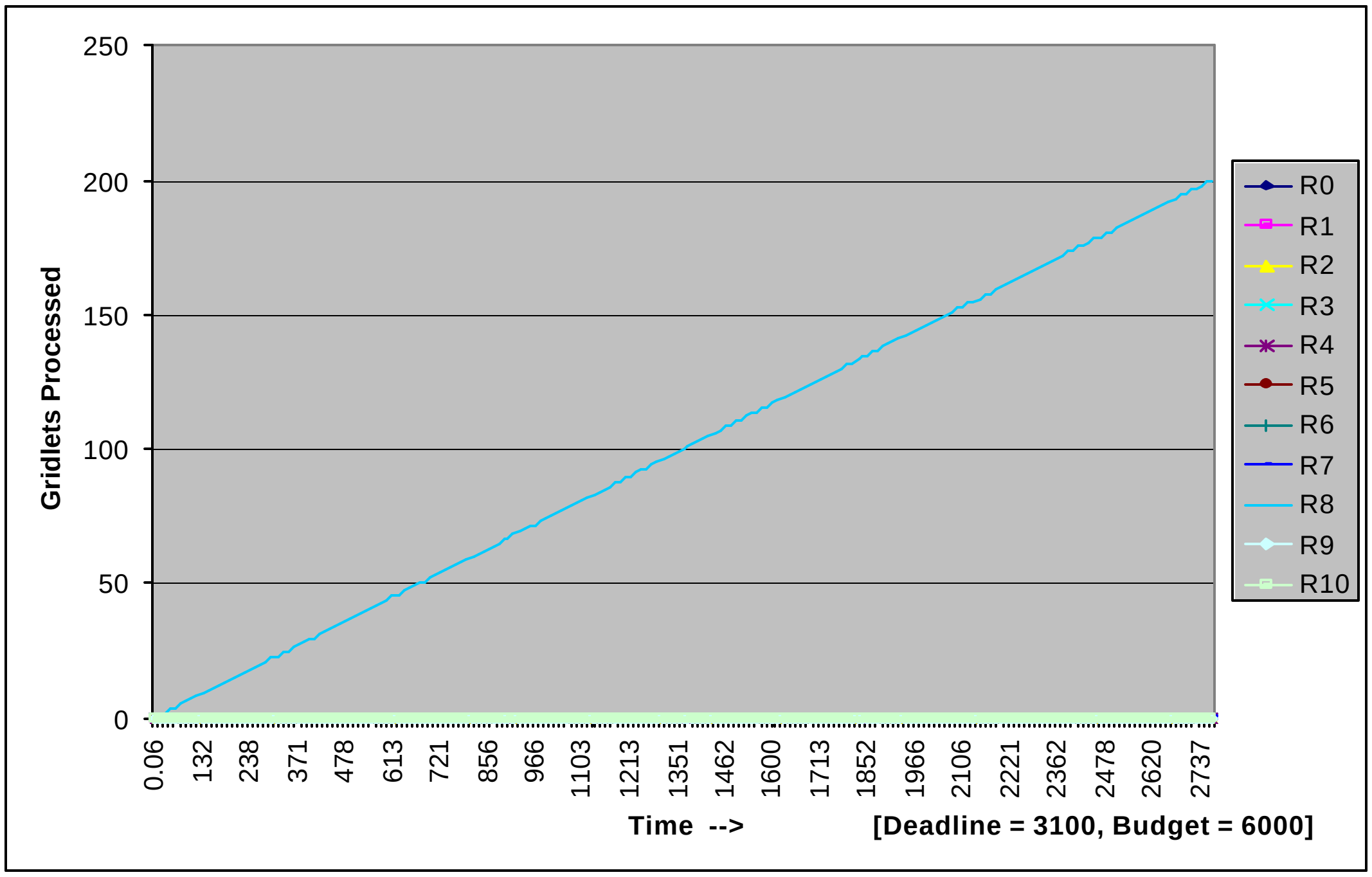

**Figure 30: Trace of No. of Gridlets processed for a high deadline and low budget constraints.**

A trace of the number of Gridlets committed to resources at different times depending on their performance, cost, and the user constraints (deadline and budget) and requirements (cost-optimization) is shown in Figure 31 and Figure 32 for deadline values of 100 and 1100 time units respectively. In both graphs it can be observed the broker committed Gridlets to expensive resources only when it is required. It committed as many Gridlets as the cheaper resources can consume by the deadline. The remaining Gridlets are assigned to expensive resources. The broker used expensive resources in the beginning and continued to use cheaper resources until the end of the experiment. This ability of economic grid broker to select resources dynamically at runtime demonstrates its adaptive capability driven by the user's quality of service requirements.

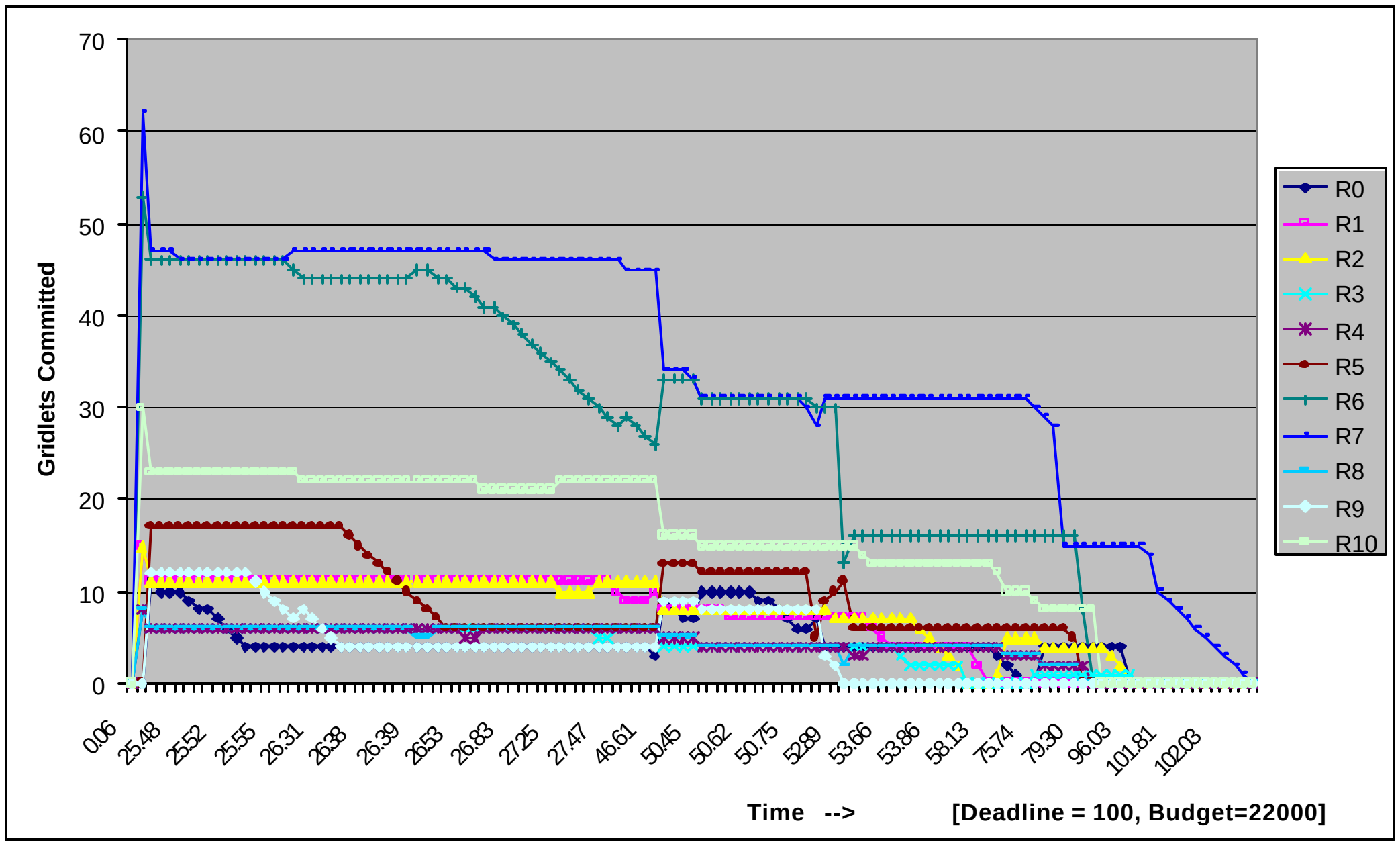

**Figure 31: Trace of the number of Gridlets committed to resources for a low deadline and high budget constraints.**

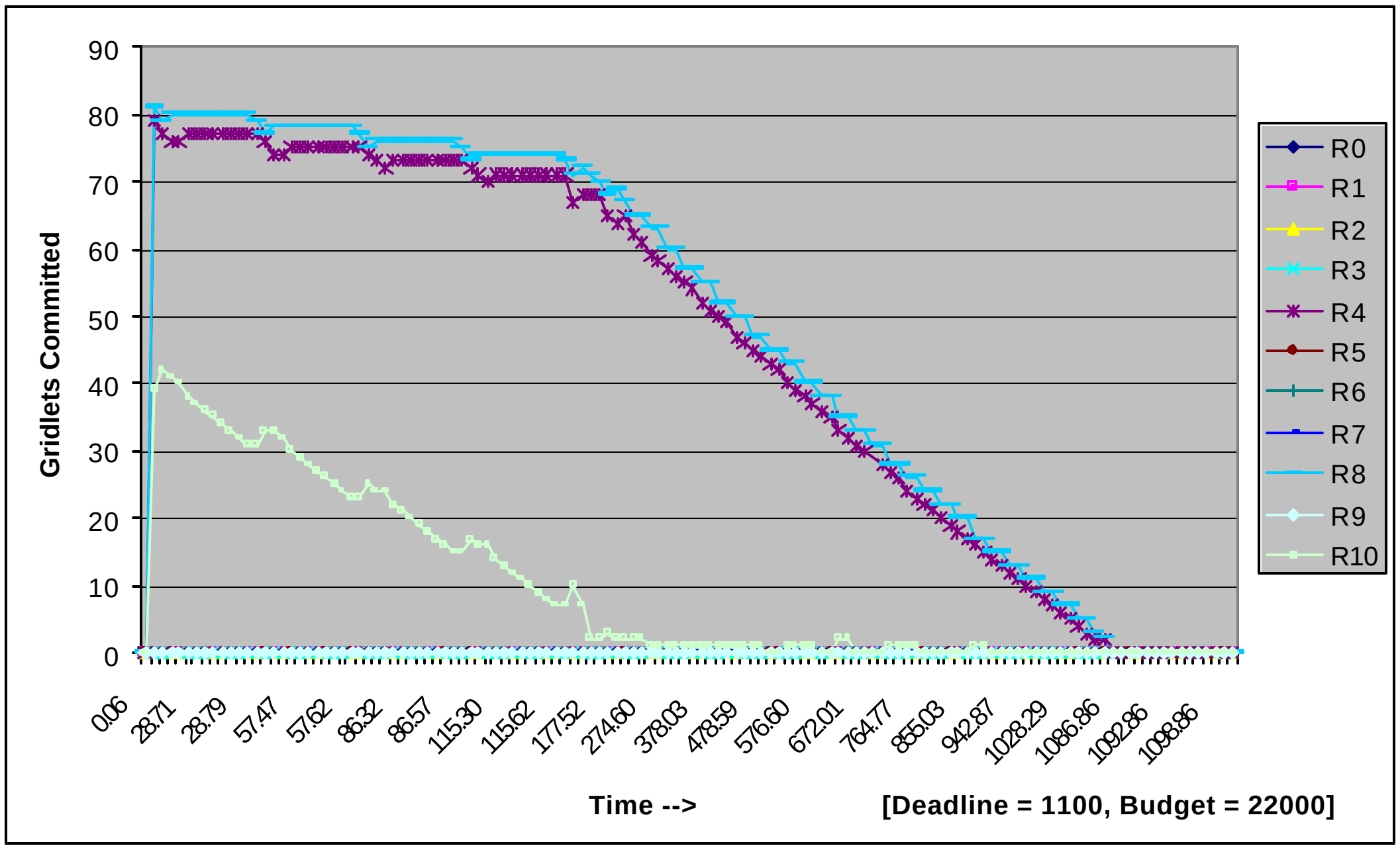

**Figure 32: Trace of the number of Gridlets committed to resources for a medium deadline and high budget constraints.**

### 5.4 DBC Scheduling Experiments with Cost-Optimization—for Multiple Users

In the second experiment, we explored distributed economic scheduling for a varying number of users competing for the same set of resources. All users are modeled to have similar requirements to enable comparison among them and understand the overall scenario. Each user application contains 200 Gridlets with small variation as explained in application modeling section. We modeled varying number of users in series from 1, 10, 20, and so on up to 100 and each with their own broker scheduling Gridlets on simulated WWG testbed resources (listed in Table 2). We explored scheduling of Gridlets for different budget values varied from 5000 to 22000 in step of 1000. For this scenario, we performed two scheduling experiments with two different values of deadline for DBC constrained *cost minimization* algorithm.

#### 5.4.1 User Deadline = 3100 time unit

The number of Gridlets processed, average time at which simulation is stopped, and budget spent for different scheduling scenario for each user with the deadline constraint of 3100 time unit is shown in Figure 33, Figure 34, and Figure 35. From Figure 33, it can be observed that as the number of users competing for the same set of resources increase, the number of Gridlets processed for each user is decreasing because they have tight deadline. Whether there are few users (e.g., 1 or 10 users in this case), they are able to process all jobs in most cases when the budget is increased. Figure 34 shows the time at which broker terminated processing of Gridlets. When a large number of users are competing (e.g., 100) for resources, it can be observed that the broker exceeded the deadline. Because, the broker initially planned scheduling Gridlets for the period of deadline, but that schedule had to be terminated because competing users had already occupied high resource share well before the recalibration phase (the first establishment of the amount of resource share available to the user, which of course can change). Figure 35 shows the average budget spent by each user for processing Gridlets shown in Figure 33, which is also clear from the graphic similarity between both diagrams when a large number of users are competing for resources.

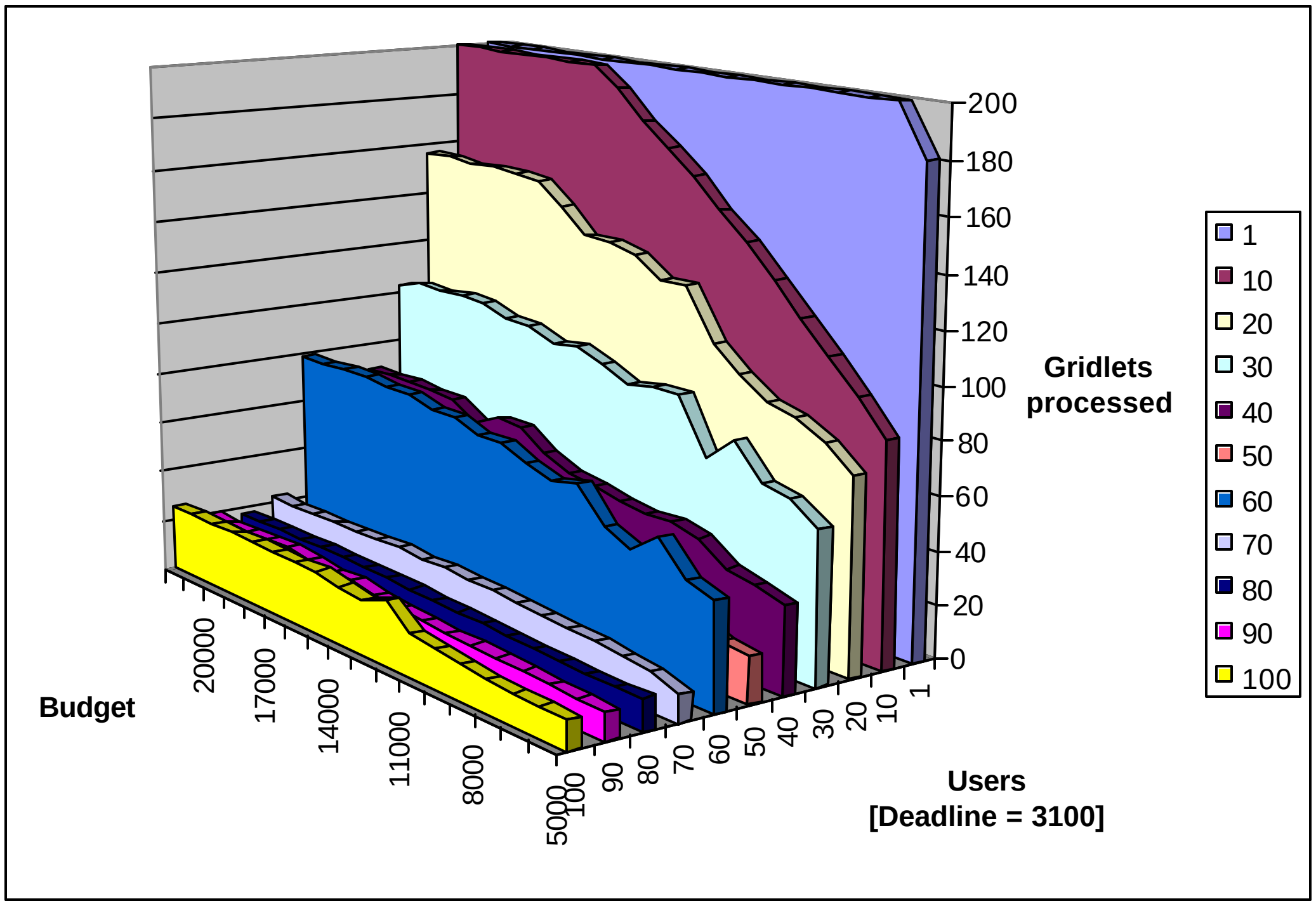

**Figure 33: No. of Gridlets processed for each user when a varying number of users competing for resources.**

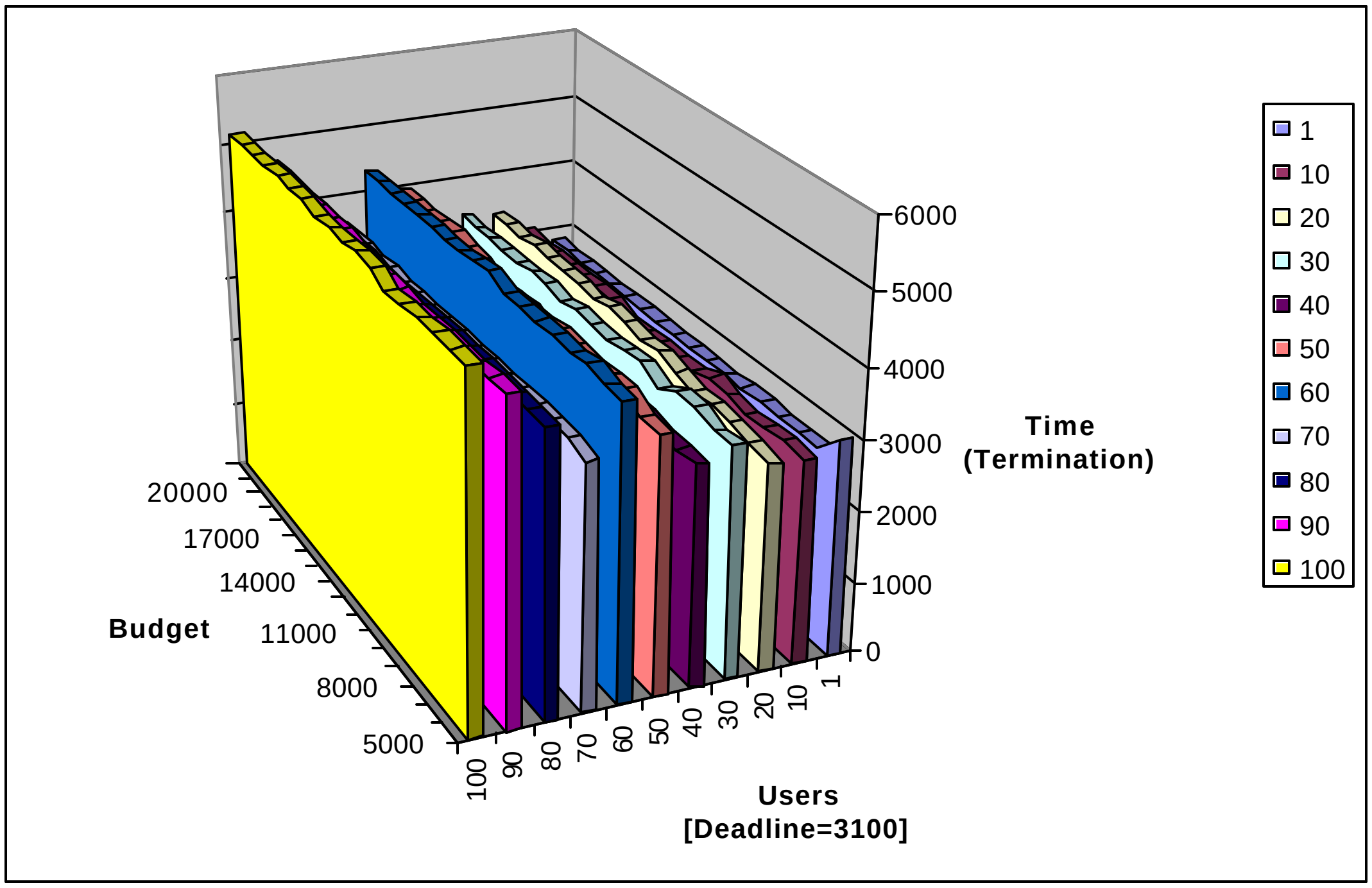

**Figure 34: The average time at which the user experiment is terminated with varying number of users competing for resources. When there are a large number of users arriving at different times, they are likely to impact on the schedule and the execution time of jobs already deployed on resources. The broker waiting for the return of jobs that are deployed on resources leads to the termination time exceeding the soft deadline unless the execution of jobs is cancelled immediately.**

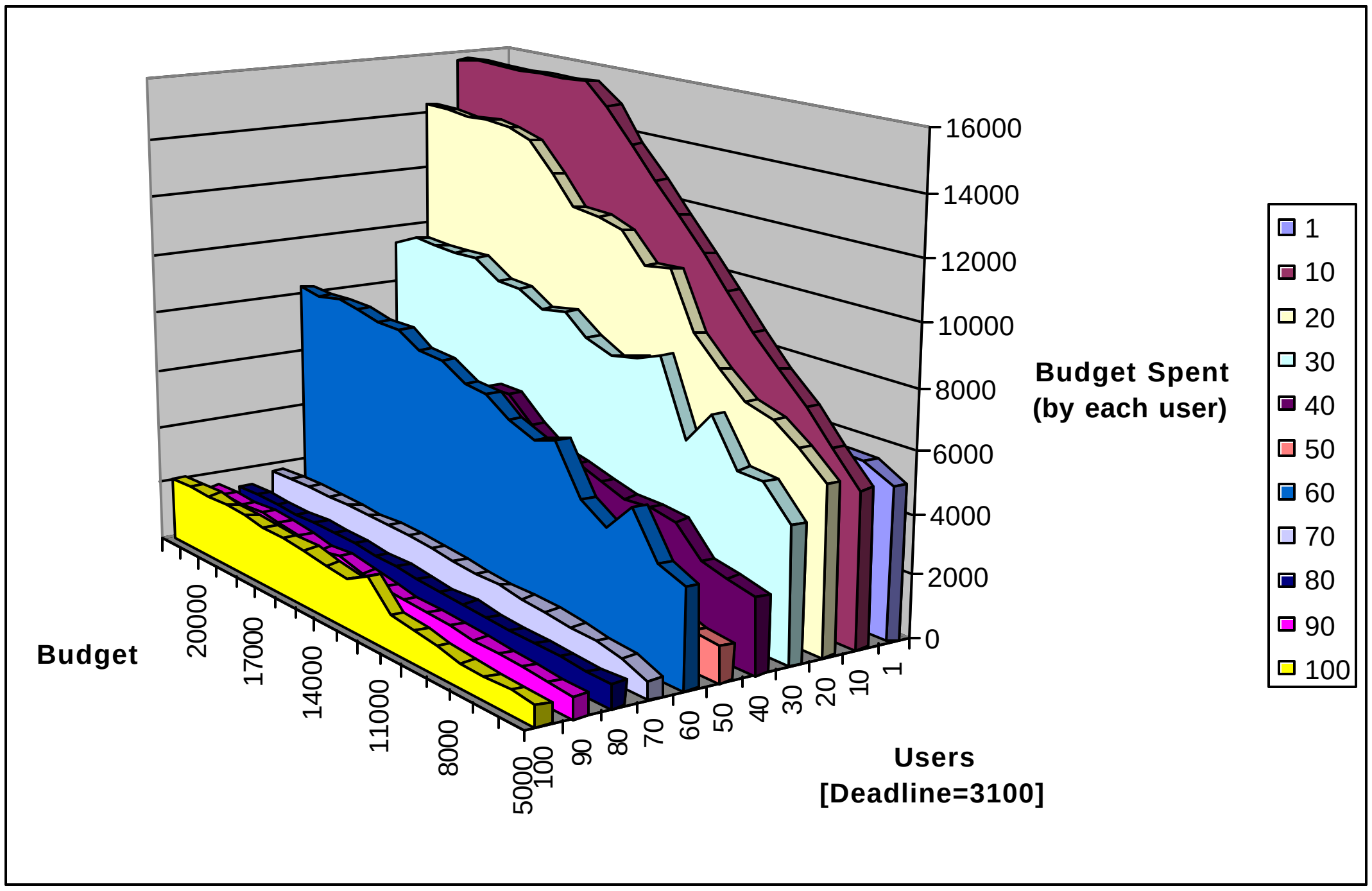

**Figure 35: The average budget spent by each user for processing Gridlets.**

### 5.4.2 User Deadline = 10000 time unit

The number of Gridlets processed, average time at which simulation is stopped, and budget spent for different scheduling scenario for each user with the deadline constraint of 10000 time unit is shown in Figure 36, Figure 37, and Figure 38. In this experiment, the number of Gridlets processed for each user improved substantially due to the relaxed deadline constraint compared to the previous experiment (see Figure 33 and Figure 36). As the number of users competing for resources increased, the number of Gridlets processed for each user decreased. But when the budget is increased, the number of Gridlets processed increased. Unlike the previous experiment, the broker is able to learn and make better predictions on the availability of resource share and the number of Gridlets that can be finished by deadline. As the deadline was sufficient enough to revisit the past scheduling decisions, the broker is able to ensure that the experiment is terminated within the deadline for most of the time (see Figure 37). The average budget spent by each user for processing Gridlets is shown in Figure 38, which is also clear from the graphic similarity between Figure 36 and Figure 38 when a large number of users are competing for resources.

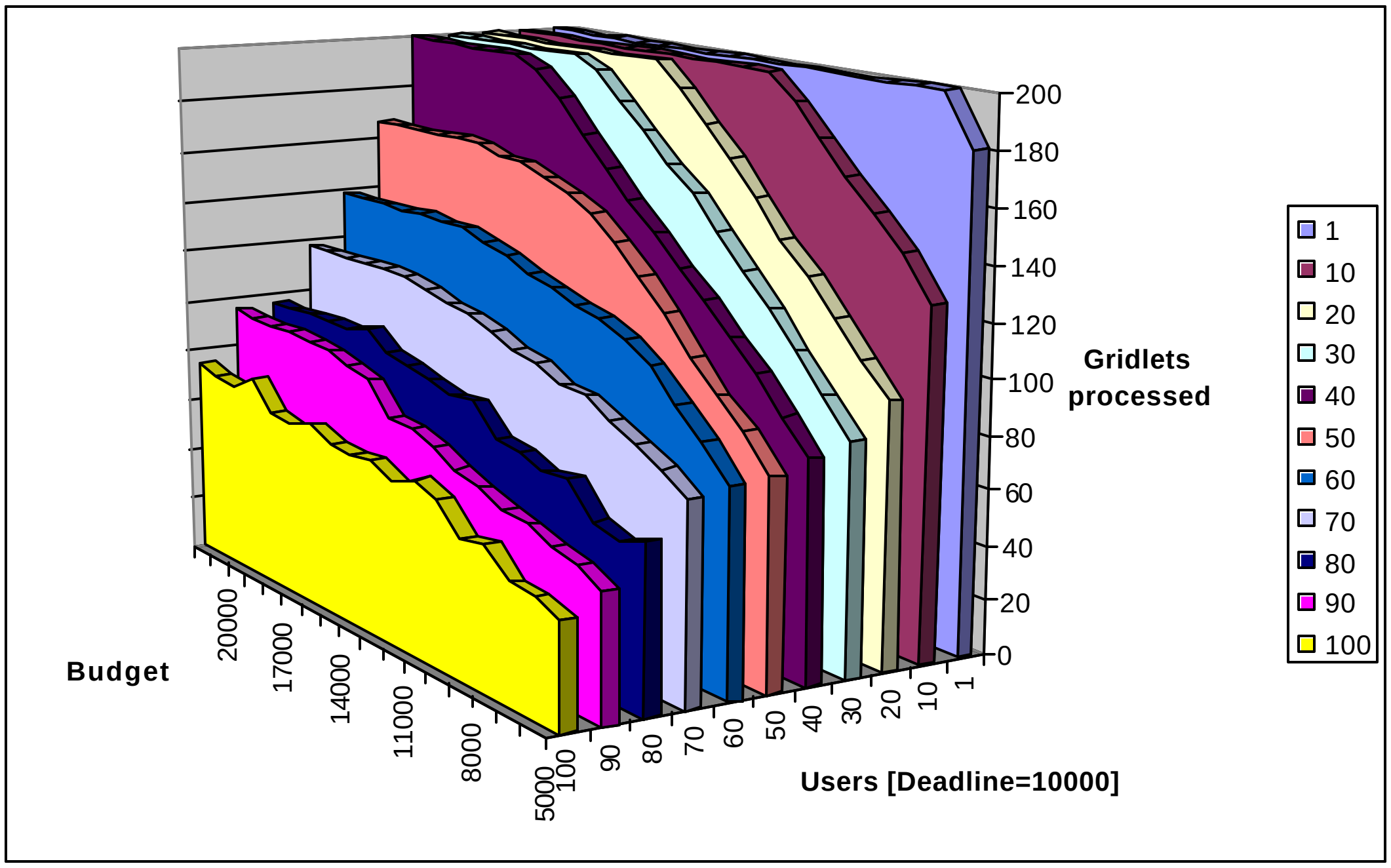

**Figure 36: No. of Gridlets processed for each user with varying number of users competing for resources.**

## 6 Conclusion and Future Works

We discussed an object-oriented toolkit, called GridSim, for resource modeling and scheduling simulation. GridSim simulates time- and space-shared resources with different capabilities, time zones, and configurations. It supports different application models that can be mapped to resources for execution by developing simulated application schedulers. We have discussed the architecture and components of the GridSim toolkit along with steps involved in creating GridSim based application-scheduling simulators.

The implementation of GridSim toolkit in Java is an important contribution since Java provides a rich set of tools that enhance programming productivity, application portability, and a scalable runtime environment. As the JVM (Java Virtual Machine) is available for single, multiprocessor shared or distributed machines such as clusters, GridSim scales with them due to its concurrent implementation. Also, we were able to leverage the existing basic discrete-event infrastructure from SimJava while implementing the GridSim toolkit.

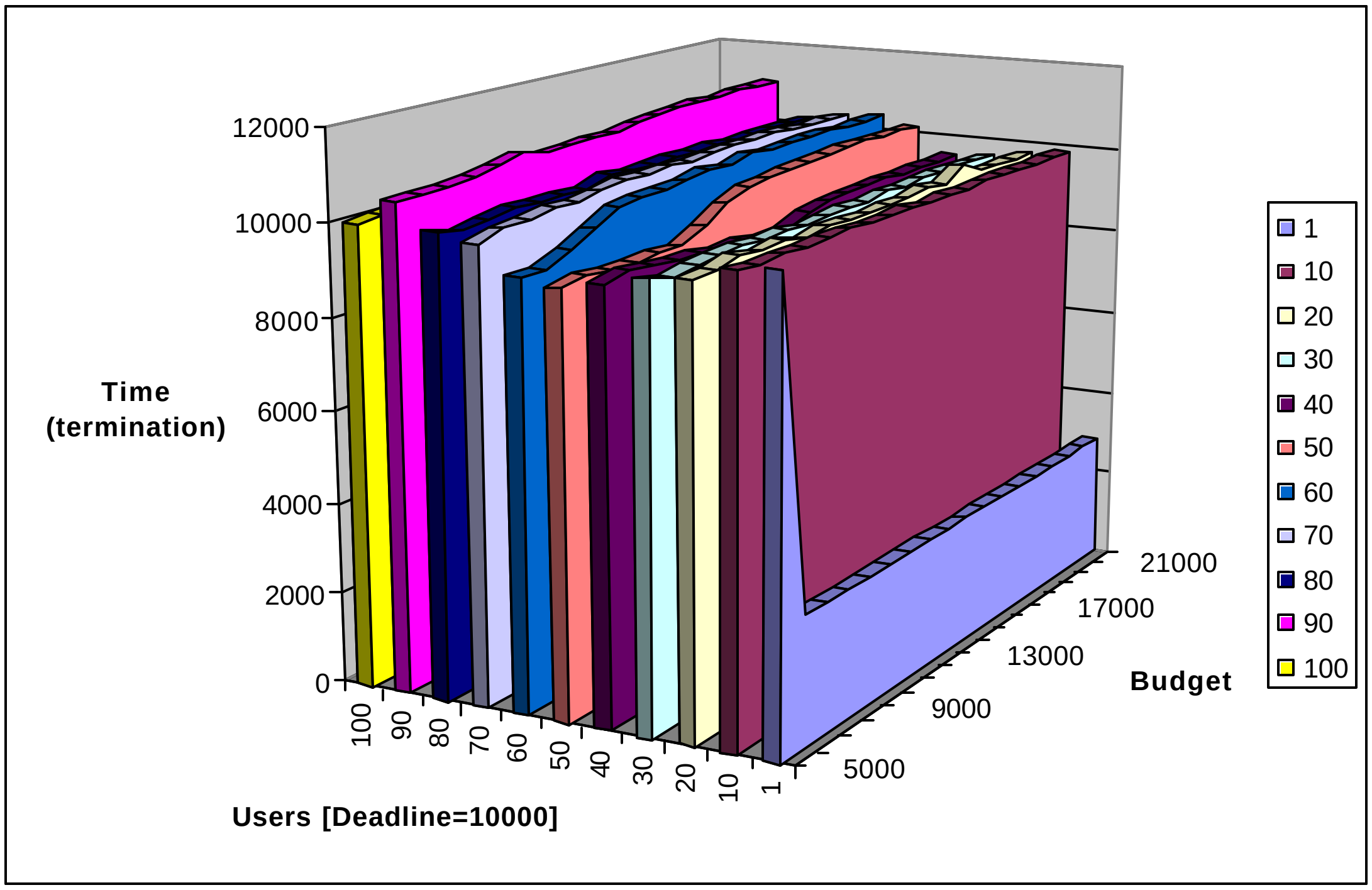

**Figure 37: The average time at which the user experiment is terminated with varying number of users competing for resources.**

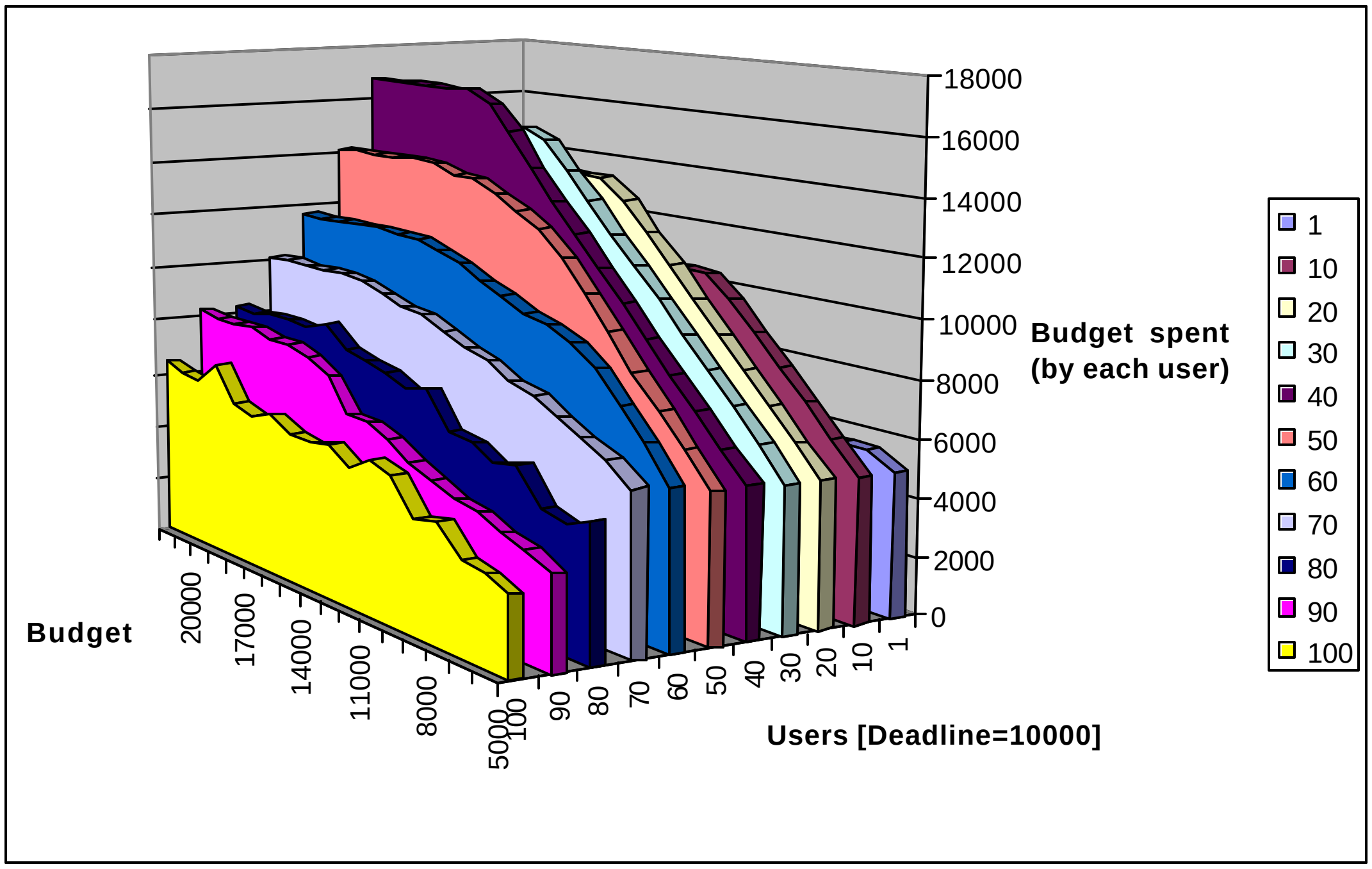

**Figure 38: The average budget spent by each user for processing Gridlets.**

We have developed a Nimrod-G like economic grid resource broker simulator using GridSim and evaluated a number of scheduling algorithms based on deadline and budget based constraints. This helped us in evaluating performance and scalability of our scheduling policies with different grid configurations such as varying number of resources, capability, cost, users, and processing requirements. The results are promising and demonstrate the suitability of GridSim for developing simulators for scheduling in parallel and distributed systems.

A better network model is required to support the application model with tasks collaborating and exchanging partial results among themselves in a peer-to-peer fashion. The future work focuses on strengthening the network model by supporting various types of networks with different static and dynamic configurations and cost-based quality of services. The resource models need to be enhanced by interfacing with off-the-shelf storage I/O simulators. We have created a resource model for advance reservation and we will be implementing its scheduling simulation. To enable simulation of grid resource management and scheduling with economic models such as tenders and auctions [24], we plan to integrate or support the FIPA (Foundation for Intelligent Physical Agents) standards [21] based interaction protocol infrastructure and extend the resource model to support them, along with quality of service guarantees. Efforts are currently underway, to develop and simulate an economic-based scheduler for single administrative domain resources such as clusters.

## Software Availability

The GridSim toolkit software with source code can be downloaded from the project website:

`http://www.buyya.com/gridsim/`

## Acknowledgements

We would like to thank David Abramson for his support and comments on improving the work. We thank John Crossley and Rob Gray for proofreading the paper. We thank Ajith Abraham and his family for their hospitality during Rajkumar's visit to the Gippsland campus! We thank Marcelo Pasin for his help in modeling resources with SPEC benchmark ratings.